\documentclass[rmp,a4paper,twocolumn,byrevtex,floatfix]{revtex4}
\usepackage{latexsym, textcomp}
\usepackage{amssymb,amsmath,longtable,bm,color}
\usepackage[Euler]{upgreek}
\usepackage{graphicx}
\usepackage{longtable}
\usepackage{mykeys}
\bibpunct[, ]{[}{]}{,}{s}{,}{,}

\graphicspath{{figsdata/}}

\begin{document}


\title{Gaussian approximation for finitely extensible bead-spring chains with hydrodynamic interaction}
\date{\today}
\author{R. Prabhakar}
\thanks{Present address: Research School of Chemistry, Australian National University, ACT - 0200, Australia}
\affiliation{Department of Chemical Engineering,
\\Monash University, Clayton, Victoria - 3800,
\\Australia}
\author{J. Ravi Prakash}
\thanks{Corresponding author}
\affiliation{Department of Chemical Engineering,
\\Monash University, Clayton, Victoria - 3800,
\\Australia}

\begin{abstract}
The Gaussian Approximation, proposed originally by \"{O}ttinger
[\emph{J. Chem. Phys.}, 90 (1) : 463-473, 1989] to account for the
influence of fluctuations in hydrodynamic interactions in Rouse
chains, is adapted here to derive a new mean-field approximation
for the FENE spring force. This "FENE-PG" force law approximately
accounts for spring-force fluctuations, which are neglected in the
widely used FENE-P approximation. The Gaussian Approximation for
hydrodynamic interactions is combined with the FENE-P and FENE-PG
spring force approximations to obtain approximate models for
finitely-extensible bead-spring chains with hydrodynamic
interactions. The closed set of ODE's governing the evolution of
the second-moments of the configurational probability distribution
in the approximate models are used to generate predictions of
rheological properties in steady and unsteady shear and uniaxial
extensional flows, which are found to be in good agreement with
the exact results obtained with Brownian dynamics simulations. In
particular, predictions of coil-stretch hysteresis are in
quantitative agreement with simulations' results. Additional
simplifying diagonalization-of-normal-modes assumptions are found
to lead to considerable savings in computation time, without
significant loss in accuracy.
\end{abstract}

\maketitle

\section{\label{s:intro} Introduction}

The importance of hydrodynamic interactions (HI) in determining
the behaviour of dilute polymer solutions has been widely
recognized since the introduction of the Zimm
theory~\citep{zimm:jcp-56}. The limitation, however, of the
accuracy of Zimm theory to linear viscoelastic property
predictions has spurred the development of more sophisticated
treatments of HI. For instance, while the consistent averaging
approximation~\citep{ottinger:jcp-87c} replaces the equilibrium
averaged Oseen tensor in Zimm theory with a non-equilibrium
average, the Gaussian approximation~\citep{ottinger:jcp-89,
wedgewood:jnnfm-89} further accounts for fluctuations in HI. The
Gaussian approximation, which is the most sophisticated
approximation to date, has been shown to be highly accurate in
shear flows by comparing its predictions of viscometric functions
with exact Brownian dynamics simulations
(BDS)~\citep{zylka:jcp-91}.

The use of bead-spring chain models with linear Hookean springs in
all these models has prevented the evaluation of their accuracy in
extensional flows, in which the polymer coil undergoes a
coil-stretch transition at a critical value $\Wi_\textrm{c}$ of
the Weissenberg number $W\!i = \lambda \, {\dot \epsilon}$, where
$\lambda$ is a characteristic relaxation time, and ${\dot
\epsilon}$ is the extension rate. As is well known, the
extensional viscosity becomes unbounded at $W\!i_\text{c}$ because
of the infinite extensibility of Hookean springs. The prediction
of a bounded extensional viscosity was achieved in early theories
by replacing Hookean springs with nonlinear finitely extensible
springs~\citep{peterlin:jpsb-66}. However, these theories did not
incorporate HI. The important role played by HI in extensional
flows has been established beyond doubt in a series of recent
publications, which have shown that the inclusion of HI leads to
quantitatively accurate predictions of both macroscopic and
mesoscopic properties~\citep{jendrejack:jcp-02, hsieh:jnnfm-03,
prabhakar:jor-04, sunthar:macromol-05a, sunthar:macromol-05b}. All
these theoretical predictions have been obtained by carrying out
exact BDS incorporating fluctuating HI and nonlinear finitely
extensible springs.

An additional aspect that has been shown in these works to be
crucial to obtaining quantitative predictions is the need to use
bead-spring chain models with a sufficiently large number of
springs, $\NS$ (or equivalently, degrees of freedom). This
requirement is eventuated by the need to accurately reflect both
the changing nature of the drag experienced by the polymer as it
unravels in an extensional flow, and the self-similar character of
a high molecular weight polymer (which is responsible for the
observation of universal behaviour in dilute polymer solutions).
The requirement of large $\NS$, coupled with the need to include
HI (which leads to the CPU time scaling as $\NS^{4.5}$), makes the
use of BDS for routine calculations highly impractical. While the
development of schemes for the accelerated calculation of HI has
alleviated this problem somewhat~\citep{jendrejack:jcp-00}, there
is a pressing need for the development of approximations capable
of describing extensional flows, that are both accurate and
computationally inexpensive. The aim of this work is to introduce
such an approximation by extending previous approximations for HI,
which were based on bead-spring chain models with Hookean springs,
to models with nonlinear finitely extensible springs.

Wedgewood and \"{O}ttinger\citep{wedgewood:jnnfm-88} have
previously introduced a similar approximation by combining the
consistent averaging scheme for HI with a FENE-P approximation for
the springs. While they carried out a detailed investigation of
the predictions of the model in shear flow, they did not evaluate
its accuracy by comparison with exact BDS, nor did they obtain any
predictions in extensional flows. In this work, we introduce an
approximation that extends the previously established Gaussian
approximation by accounting for fluctuations in \emph{both} HI and
the spring forces. The accuracy of this model is then verified by
comparison with BDS in a number of both transient and steady,
shear and extensional flows.

The reduction in computational time achieved by the introduction
of such approximations, despite being significant, is not
sufficient to explore the behaviour of truly long chains necessary
to adequately model high molecular weight polymers. Several
workers have shown that mapping the bead-connector vectors to
`normal' coordinates, followed by an assumption that certain
matrices in the transformed space are diagonal, leads to further
significant reductions in CPU time, without any sacrifice in
accuracy~\citep{kisbaugh:jnnfm-90, prakash:macromol-99}. Here, we
carry out a similar diagonalization procedure, and examine both
the reduction in CPU time, and its effect on the accuracy of the
model.

The paper is organized as follows. In the following section, we
recall the basic equations for non-free-draining bead-spring
models of dilute solutions of finitely-extensible polymer
molecules.  Subsequently, we develop in section \ref{s:cla}, the
equations for the Gaussian approximation, and its diagonalized
version. Section \ref{s:mf} introduces the flows and material
functions of interest in this paper. The symmetries in the flow
patterns considered lead to considerable simplifications in the
implementation of the numerical methods for computing the material
functions, which are briefly discussed in section \ref{s:nm}.
Predictions of the approximations for the material functions in
shear and extensional flows are presented in section \ref{s:randd}
along with the exact results obtained with BD simulations. Section
\ref{s:concl} summarizes the central conclusions of this study.

\section{\label{s:basic}Basic equations}
Under homogeneous conditions, the configurational state of a
bead-spring chain, with $\NS$ springs, is completely specified by
the set of connector vectors $\{\bQ_i \,|\, i = 1,\ldots,\NS\}$
\citep{bird:dpl2}. The Fokker-Planck equation governing the
evolution of the configurational probability distribution
$\psi(\bQ_1,\ldots,\bQ_{\NS}, t)$ is,
\begin{align}
\begin{split}
\frac{\partial \psi}{\partial t} = &- \sum_{i =1}^{\NS}
\frac{\partial}{\partial \bQ_i} \,\cdot\, \left\{\bk \cdot \bQ_i -
\frac{1}{\zeta} \sum_{j = 1}^{\NS} \bA_{ij} \cdot
\bm{F}_j^{\textsc{s},\,c} \right\} \psi \\&+ \frac{\kbT}{\zeta}
\sum_{i,\,j =1}^{\NS} \frac{\partial}{\partial \bQ_i} \cdot \bA_{i
j} \cdot \frac{\partial \psi}{\partial \bQ_j}\,,
\end{split}\label{e:FPEconn}
\end{align}
where, $\kb$ is Boltzmann's constant, $\bk$ is the transpose of
the position-independent traceless velocity gradient $\nabla
\bm{v}$, $T$ is the absolute temperature of the solution, and
$\zeta$ is the Stokesian bead-friction coefficient, which is
related to the bead radius $a$ through $\zeta = 6 \pi a
\eta_\text{s}$, with $\eta_\text{s}$ being the viscosity of the
Newtonian solvent. The dimensionless diffusion tensors $\bA_{i j}$
are discussed in detail shortly. The spring connector force
$\bm{F}_j^{\textsc{s},\,c}$ is given by,
\begin{gather}
\bm{F}_j^{\textsc{s},\,c} = \frac{\partial S_j}{\partial \bQ_j}\,,
\label{e:sprF}
\end{gather}
with $S_j$ representing the potential energy of spring $j$. Here,
the following FENE expression for the spring force is used
\citep{warner:iecf-72}:
\begin{gather}
\bm{F}^{\textsc{s},\,c}_j \,= \,H\,\left( \frac{1}{1 -
\bQ^{2}_j/Q_0^2}\right)\, \bQ_j \,=\, H\, \xi (Q_j)\,\bQ_j \,,
\label{e:FENE}
\end{gather}
where, $H$ is the spring constant, $Q_j = | \,\bQ_j\, |$, $Q_0$ is
the `fully-stretched' length a single spring, and $\xi= \left[ {1
- \bQ^{2}_j/Q_0^2}\right]^{-1}$ is the nonlinearity in the FENE
spring force law.

By considering a bead-spring chain of $\NS$ springs to represent a
polymer molecule of $\NK$ Kuhn-segments, each spring can be
imagined to represent a ``sub-molecule" of $\NKS = \NK/\NS$ Kuhn
segments. The parameters in the FENE spring-force law, $H$ and
$Q_0$, can then be related to the two parameters characterizing
the equilibrium structure of a sub-molecule, namely, $\bK$ (the
length of a Kuhn step) and $\NKS$. Since the contour length of the
sub-molecule is $\bK \NKS$, it follows that $Q_0 = \bK \NKS$. The
equilibrium mean square end-to-end distance of a sub-molecule, in
the absence of excluded volume forces, is $R_{\textsc{s}}^2 =
\bK^2 \NKS$. Using the equilibrium Boltzmann distribution for a
single spring governed by the FENE spring force law,
$\psi_\textrm{eq}^\textsc{s} \sim \exp (- S_j / \kbT)$,  it can be
shown that~\citep{bird:dpl2},
\begin{gather}
\frac{H}{\kbT} = \frac{3}{R_\textsc{s}^2} - \frac{5}{Q_0^2} \, .
\label{e:HRs}
\end{gather}
Substituting for $R_\textsc{s}^2$ and $Q_0^2$ in terms of $\bK$
and $\NKS$, leads to the following expression for the spring
contant $H$,
\begin{gather}
\frac{H}{\kbT} = \frac{1}{\bK^2}\, \frac{3 \NKS - 5}{\NKS^2} \, .
\label{e:Hdef}
\end{gather}

The equations above indicate that taking the limit $Q_0
\rightarrow \infty$, while keeping $R_\textsc{s}^2$ constant,
leads to $\xi \rightarrow 1$. In this limit, $R_{\textsc{s}}^2 = 3
\kbT / H$, and the FENE spring-force expression reduces to the
Hookean spring force law, $\bm{F}^{\textsc{s},\,c} = H \bQ$. In
this work, chains with Hookean springs are referred to as ``Rouse"
chains, while finitely-extensible bead-spring chains are denoted
as ``FEBS" chains.

The diffusion tensors $\bA_{i j}$ in Eq.~\eqref{e:FPEconn} are
given by,
\begin{gather}
\bA_{ij} = A_{ij} \ut + \zeta \left( \bO_{ij} + \bO_{i+1,\,j+1} -
\bO_{i,\,j+1} - \bO_{i+1,\,j} \right)\,, \label{e:bAdef}
\end{gather}
where, $A_{ij} = \delta_{ij} + \delta_{i+1,\,j+1} -
\delta_{i,\,j+1} - \delta_{i+1,\,j} = 2 \delta_{ij} -
\delta_{|i-j|,1}$ is an element of the $\NS \times \NS$ Rouse
matrix, and $\bO_{\nu\mu}$, $\nu, \mu = 1,\ldots,N$, are HI
tensors, which are typically related to the inter-bead
displacement $\br_{\mu \nu}$ between beads $\mu$ and $\nu$, by
expressions of the form,
\begin{gather}
\bO_{\nu \mu} = \frac{1}{8 \pi \etas \, r_{\mu \nu}}\,
\bm{C}(\br_{\mu \nu}) \, . \label{e:HIT}
\end{gather}
Note that the inter-bead displacement vector $\br_{\mu \nu}$ can
be related to the spring connector vectors through,
\begin{equation}
\br_{\mu \nu} = \textrm{sgn} (\nu - \mu) \sum_{i \,=\, \min(\nu,
\mu)}^{\max (\nu, \mu) - 1} \bQ_i  \, . \label{e:rnumuvsQ1}
\end{equation}

Models which neglect the influence of HI---such as the original
bead-spring model of Rouse \citep{rouse:jcp-53}---can be
interpreted as ones in which the tensor $\bm{C}$ is set to zero.
Early models incorporating HI \citep{kirkwood:jcp-48, zimm:jcp-56}
used the Oseen-Burgers form of the HI tensor, in which
\begin{gather}
\bm{C} = \ut + \frac{\br_{\mu\nu} \br_{\mu\nu}}{r_{\mu \nu}^2}
\label{e:OBt}.
\end{gather}
Although the Oseen-Burgers description of HI is inaccurate when
the pair-wise separation between beads is comparable to the size
of the beads, the hydrodynamic behaviour of long polymer chains is
dominated by the interactions between non-adjacent beads, and the
short-range inaccuracy of the Oseen-Burgers description has been
shown to be relatively unimportant in the prediction of properties
that depend on the polymer chain as a whole
\citep{ottinger:jcp-87c, zimm:macromol-80}.

Further, its simple form is particularly useful for the
development of closure approximations. In numerical simulations,
however, the Oseen-Burgers expression becomes problematic when
$r_{\nu \mu} \leq a$, since the  $3N \times 3N$ diffusion
block-matrix comprising the $\bA_{i j}$ tensors as its constituent
blocks, can become non-positive definite \citep{rotne:jcp-69} when
beads overlap. In its place, in all the Brownian dynamics
simulations performed here, the Rotne-Prager-Yamakawa (RPY)
modification \citep{rotne:jcp-69, yamakawa:mtps},
\begin{gather} \bm{C} = \begin{cases} \left(1 +
\displaystyle{\frac{2}{3} \frac{a^2}{r_{\mu \nu}^2}}  \right) \ut
+ \left(1- \displaystyle{\frac{2 a^2}{r_{\mu \nu}^2}} \right)
\displaystyle{\frac{\br_{\mu\nu} \br_{\mu\nu}}{r_{\mu \nu}^2}}\,,
r_{\mu \nu} \geq 2a,  \\[1cm]
\displaystyle{\frac{r_{\mu \nu}}{ 2 a}} \left[ \left(
\displaystyle{\frac{8}{3} - \frac{3 r_{\mu \nu}}{4 a} }\right)\ut
+ \displaystyle{\left(\frac{r_{\mu \nu}}{4 a}\right)
\frac{\br_{\mu\nu} \br_{\mu\nu}}{r_{\mu \nu}^2}} \right]\,, r_{\mu
\nu} < 2a. \end{cases} \label{e:RPY}
\end{gather}
which ensures that the diffusion block-matrix always remains
positive-definite, is used.

In order to express the Fokker-Planck equation in dimensionless
form, the following length scale and associated time scale are
introduced \citep{prabhakar:jor-04, sunthar:macromol-05a},
\begin{gather}
\elS \equiv \frac{R_{\textsc{s}}} {\sqrt{3}} \, ; \quad \lS \equiv
\frac{\zeta \elS^2}{4 \kbT} \,. \label{e:lS}
\end{gather}
In the Hookean limit, both $\elS$  and $\lS$ reduce to their
conventional form, $\elS = \elH = \sqrt{\kbT/H}$ and $\lS = \lH =
\zeta/ 4H = \zeta \elH^2 /4 \kbT $, respectively. Defining $\ts
\equiv t/ \lS$, $\bQs_i \equiv \bQ_i/ \elS$, $\bks \equiv \lS
\bk$, the Fokker-Planck equation can be recast in dimensionless
terms as
\begin{align}
\begin{split}
\frac{\partial \psi}{\partial \ts} = &- \sum_{i =1}^{\NS}
\frac{\partial}{\partial \bQs_i} \,\cdot\, \left\{\bks \cdot
\bQs_i - \frac{H^\ast}{4} \sum_{j = 1}^{\NS} \bA_{ij} \cdot \xi_j
\bQs_j \right\} \psi \\ &+ \frac{1}{4} \sum_{i,\,j =1}^{\NS}
\frac{\partial}{\partial \bQs_i} \cdot \bA_{i j} \cdot
\frac{\partial \psi}{\partial \bQs_j} \,, \end{split}
\label{e:FPEconnd}
\end{align}
where,
\begin{gather}
H^\ast \equiv \frac{H \elS^2}{\kbT} = \frac{\elS^2}{\elH^2}\,,
\label{e:caphsdef}
\end{gather}
is the dimensionless spring constant. When $\elH$ is used as the
length scale, the parameter $Q_0$ is represented in dimensionless
terms by the finite extensibility parameter, $b \equiv Q_0^2 /
\elH^2$ \citep{bird:dpl2}. The counterpart of this parameter in
the non-dimensionalization scheme adopted here is, $\bs \equiv
{Q_0^2}/{\elS^2}$. Since $Q_0 = \bK \NKS$, it follows that, $\bs =
3 \NKS$. Because of the direct relation of $\bs$ to the degree of
coarse-graining $\NKS$, the latter is often reported instead of
$\bs$ when presenting results.

On introducing the hydrodynamic interaction parameter, $\hs$,
defined by,
\begin{gather}
\hs = \frac{a}{\sqrt{\pi} \elS}\, , \label{e:hsdef}
\end{gather}
the dimensionless tensor $\zeta \bO_{\nu \mu}$ occurring in the
definitions of the diffusion tensors [Eq.~\eqref{e:bAdef}] can be
written as
\begin{gather}
\zeta \bO_{\nu \mu} = \frac{3\sqrt{\pi} \hs}{4 \, r_{\mu
\nu}^\ast}\, \bm{C}(\br_{\mu \nu}^\ast)\,, \label{e:HITnd}
\end{gather}
where $\br_{\mu \nu}^\ast = (1/\elS) \br_{\mu \nu}$.

All the rheological properties of interest in the present
work--which are discussed in greater detail in Section
\ref{s:mf}--can be obtained from Kramer's expression for the
polymer contribution to the stress tensor
$\btaup$~\citep{bird:dpl2}, which in non-dimensional terms is,
\begin{align}
\btaups =  \frac{1}{\np \kbT} \btaup = \NS \ut - H^\ast \sum_{i =
1}^{\NS} \langle \xi_i \bQs_i \bQs_i \rangle \,. \label{e:kramer}
\end{align}
Here, $\np$ is the number density of polymers, $\ut$ is the unit
tenor, and angular brackets denote an average performed with the
configurational distribution function $\psi$.

Besides rheological properties, the mean square end-to-end
distance $\Resqs$ is used here to provide information on the size
and shape of polymer molecules as they stretch and orient in
response to the imposition of flow,
\begin{gather}
\Resqs = \sum_{i,j = 1}^{\NS}  \tr \QQs{i}{j}\,. \label{e:resqs}
\end{gather}
Further, the prediction of the various models of the birefringence
in extensional flows is evaluated using the following expression
\citep{wiest:polymer-99, doi:tpd}:
\begin{gather}
\Delta n^\ast \equiv \frac{n^{\textsc{i}}_{xx} - n^{\textsc{i
}}_{yy}}{A}\, = \frac{1}{\bs} \sum_{i = 1}^{\NS} [\langle
{\Qs}^2_{i, \,xx}  \rangle - \langle {\Qs}^2_{i,\,yy}  \rangle]\,.
\label{e:dn*}
\end{gather}
where, $\mathbf{n}^\textsc{i}$ is the intrinsic birefringence
tensor, and $A$ is a constant that, besides depending on the
chemical structure of the polymer and the solvent, is proportional
to the polymer concentration $\np$.

Equations~\ref{e:kramer}, \ref{e:resqs} and \ref{e:dn*} above show
that all the macroscopic properties of interest here can be
described in terms of expectations of configuration dependent
functions. It is possible to obtain formally the equation for the
time evolution of any configuration dependent function by
multiplying the Fokker-Planck equation with the function and then
integrating over all possible configurations. However, as a direct
consequence of the nonlinearities in the model arising from FE and
HI, the resulting equations are not closed with respect to the
expectations whose evolution equations are desired. For instance,
the following set of evolution equations for the second moments of
the probability distribution $\psi$ can be derived in this manner:
\begin{align}
\QQs{i}{j}_{(1)} &\equiv \frac{d \QQs{i}{j}}{d \ts} - \bks \cdot
\QQs{i}{j} - \QQs{i}{j} \cdot {\bks}^{\textsc{t}} \notag \\&= -
\frac{\Hs}{4} \sum_{m=1}^{\NS} \langle \bQs_i \bQs_m \xi_m \cdot
\bA_{mj} + \bA_{im} \cdot \xi_m \bQs_m \bQs_j \rangle \notag \\
& \quad + \frac{1}{2} \langle  \bA_{ij} \rangle\,, \label{e:xqq}
\end{align}
where the subscript ``(1)" denotes the contravariant convected
time derivative of a tensor \citep{bird:dpl1}. As a result of the
nonlinearities in the functions $\xi$ and $\bO$ introduced by the
spring force law and HI respectively, the equation above for
$\QQs{i}{j}$ is not closed with respect to the second moments. It
is this `closure problem' that has motivated both the introduction
of various closure schemes, and the development of exact Brownian
dynamics simulations for obtaining predictions of properties away
from equilibrium. As mentioned previously, the Gaussian
approximation for Rouse chains is the most accurate closure
approximation to date. In the section below, we develop the
Gaussian approximation for FEBS chains with HI.

\section{\label{s:cla}Gaussian approximation}

For Rouse chains, \"{O}ttinger\citep{ottinger:jcp-89} and
Wedgewood\citep{wedgewood:jnnfm-89} pointed out that the
complicated averages on the right-hand side of
equation~\ref{e:xqq} above (with $\xi_m = 1$) can be expressed as
functions of the second-moments $\QQs{i}{j}$ by \emph{assuming}
\emph{a priori} that the configurational distribution function is
a multivariate Gaussian,
\begin{gather}
\psi = \mathcal{N}\, \exp\, \left[ -\half \sum_{i,j = 1}^{\NS}
\bQs_i \cdot \brho^\ast_{ij} \cdot \bQs_j \right]\,,
\label{e:gauss}
\end{gather}
where $\mathcal{N}$ is a normalization factor. The tensors
$\brho^\ast_{ij}$ are related to the second moments
$\bsigs_{ij}=\QQs{i}{j}$ of the Gaussian distribution through the
following set of $\NS^2$ linear algebraic equations
\begin{gather}
\sum_{m = 1}^{\NS} \, \bsigs_{im} \cdot \brho^\ast_{mj} =
\delta_{ij} \ut\,. \label{e:sigro}
\end{gather}
As is well known, this implies that the Gaussian distribution
$\psi$ is completely specified once its second moments
$\bsigs_{ij}$ are known.

The crucial property of Gaussian distributions that enables the
simplification of complex moments is described by Wick's
decomposition rule. In the present instance, because the right
hand side of Eq.~\eqref{e:xqq} has complex averages of the form
$\langle \bm{x} \bm{x} f( \bm{x} ) \rangle$, where $\bm{x}$ is the
random variable of interest, and $f$ is a nonlinear function of
$\bm{x}$, Wick's decomposition rule leads to
\begin{gather}
\langle \bm{x} \bm{x} f \rangle = \langle \bm{x} \bm{x} \rangle
\langle f  \rangle + \langle \bm{x} \bm{x} \rangle \cdot \langle
\frac{ \partial f }{\partial \bm{x}} \bm{x} \rangle \,.
\label{e:genga}
\end{gather}
By applying Eq.~\ref{e:genga} to Rouse chains with HI,
\citeauthor{ottinger:jcp-89} showed that Eq.~\eqref{e:xqq} (with
$\xi_m = 1$) reduces to a closed set of equations for the
second-moments $\QQs{i}{j}$. In principle, the same approach can
be used to obtain closure of Eq.~\eqref{e:xqq} with both the
nonlinear phenomena of finite extensibility and hydrodynamic
interactions included. However, directly using the Gaussian
assumption and Wick's decomposition to simplify the expectations
on the right-hand side of Eq.~\eqref{e:xqq} leads to the incorrect
prediction of a dependence of equilibrium static properties on
$\hs$. Similar behaviour was earlier observed by Prabhakar and
Prakash\citep{prabhakar:jor-02} when the Gaussian approximation
was used to achieve closure in a Hookean dumbbell model
simultaneously incorporating excluded volume and hydrodynamic
interactions. The reason for this incorrect prediction is that, on
direct application of the Gaussian approximation, the
contributions from FE (and/or excluded volume) and those from HI
cannot be separated into independent factors at equilibrium
\citep{prabhakar:phd-05}, thus leading to an unphysical dependence
of the predicted equilibrium second moments on the HI parameter,
$\hs$. To avoid this problem, closure of Eq.~\eqref{e:xqq} is
achieved here in two steps. In the first step, a mean-field
quadratic spring potential $\overline{S}^\ast_m$, which
approximately accounts for fluctuations in the FENE spring force,
is derived by using the Gaussian approximation for
\textit{free-draining} FEBS chains. A bead-spring chain model is
then constructed, which incorporates hydrodynamic interactions,
and which has springs described by $\overline{S}^\ast_m$. In the
second step, the Gaussian approximation is again invoked to obtain
a closed set of equations for the second moments of this model. In
this manner, the incorrect prediction of a dependence of
equilibrium static properties on $\hs$ is avoided.

Before discussing these two steps in detail in
sections~\ref{s:fenepg} and~\ref{s:gafebs} below, it is
appropriate to note that if $f$ had been replaced with its average
$\langle f \rangle$ in Eq.~\ref{e:genga}, thus neglecting the
fluctuations in $f$, one obtains the first term on the right-hand
side alone. Thus, for a Gaussian distribution, it is seen that
fluctuations in $f$ are exactly accounted for by the second term
on the right-hand side of the equation above. This convenient
separation of the complicated expectation on the left hand side
into contributions from the mean and fluctuations will be used
here to highlight the important role played by fluctuations in the
spring force and in hydrodynamic interactions.

\subsection{\label{s:fenepg} Free-draining FEBS chains}

In this section, the Gaussian approximation is used to achieve
closure of the second-moment equations for chains with
finitely-extensible springs, but without HI. Prakash and
co-workers have previously used the Gaussian approximation for
bead-spring chain models without HI, but which are, however, still
non-linear because of the presence of conservative
non-hydrodynamic excluded volume forces between beads. The
approximation was found by them to lead to qualitatively accurate
predictions~\citep{prakash:macromol-99, prakash:macromol-01,
prakash:ces-01, kumar:jcp-04}.

For free-draining chains, $\bA_{ij} = A_{ij} \ut$, and hence from
Eq.~\eqref{e:xqq}, one obtains for FEBS chains
\begin{align}
\begin{split}
\QQs{i}{j}_{(1)} &= - \frac{\Hs}{4} \sum_{m=1}^{\NS} \left[\langle
\bQs_i \bQs_m \xi_m \rangle A_{mj} + A_{im} \langle \xi_m \bQs_m
\bQs_j \rangle\right] \\
&+ \frac{1}{2} A_{ij} \ut \,.
\end{split}
\label{e:xqqfe}
\end{align}
The equation for the polymer stress is the same as in
Eq.~\eqref{e:kramer}. In this case, closure of the evolution
equation above is not possible because of the FENE nonlinearity,
$\xi_m$, on the right-hand side. However, as in the case of the
Gaussian approximation for HI, assuming that the configurational
distribution is a Gaussian, and using Wicks decomposition rule,
the problematic terms on the right-hand side of the equation above
can be formally simplified. Since $\partial S^\ast_m/
\partial \bQs_m = \Hs \xi_m \bQs_m$
(where, $S^\ast_m = S_m/ \kbT$), using Eq.~\eqref{e:genga} leads
to
\begin{gather}
\Hs \langle \bQs_i \bQs_m \xi_m \rangle = \left\langle \bQs_i
\frac{\partial S^\ast_m}{\partial \bQs_m} \right \rangle =
\bsigs_{im} \cdot \left\langle \frac{
\partial^2 S^\ast_m}{\partial \bQs_m \partial \bQs_m} \right
\rangle
\end{gather}
The expression on the right-hand side could have also been
obtained by directly using a quadratic mean-field spring
potential,
\begin{gather}
\overline{S}^\ast_m \equiv \half \left\langle \frac{
\partial^2 S^\ast_m}{\partial \bQs_m \partial \bQs_m} \right
\rangle \colon \bQs_m \bQs_m \label{e:mfsprpot}
\end{gather}
in place of the exact spring potential $S^\ast_m$. In other words,
applying the Gaussian approximation to simplify the average due to
FENE springs in free-draining chains is equivalent to using the
quadratic mean-field spring potential $\overline{S}^\ast_m$, since
this approximation leads to
\begin{gather}
\left\langle \bQs_i \frac{\partial S^\ast_m}{\partial \bQs_m}
\right \rangle = \left\langle \bQs_i \frac{\partial
\overline{S}^\ast_m}{\partial \bQs_m} \right \rangle
\end{gather}
Prakash\citep{prakash:macromol-99} have previously shown that
Fixman's treatment of the excluded volume interaction
potential~\citep{fixman:jcp-66a} corresponds to the use of a
similar pair-wise mean-field quadratic excluded volume potential.

The expression for $\overline{S}^\ast_m$ in Eq.~\ref{e:mfsprpot}
can be simplified by using $\partial S^\ast_m /
\partial \bQs_m = \Hs \xi_m \bQs_m$,
\begin{gather}
\overline{S}^\ast_m  = \Hs \left[ \langle \xi_m \rangle \ut +
\left\langle \frac{\partial \xi_m}{\partial \bQs_m } \bQs_m \right
\rangle \right] \colon \bQs_m \bQs_m \,.
\end{gather}
The first term within the brackets on the right-hand side of the
equation above containing $\langle \xi_m \rangle$, the average of
the spring force nonlinearity, can be recognized as that arising
from a consistent-averaging treatment of that nonlinearity, while
the remaining term accounts for the influence of fluctuations in
the spring force.

Recursive application of Wick's decomposition rule, using the
result $\partial \xi_m/
\partial \bQs_m  = 2 \bQs_m\, d \xi_m/ d ({\Qs_m}^2)$,  leads to a series
expansion,
\begin{align}
\overline{S}^\ast_m = \Hs \left[ \langle \xi_m \rangle \ut +
\sum_{s = 1}^\infty 2^s \left\langle \frac{d^s \xi_m}{d
({\Qs}_m^{\,2})^s} \right\rangle  \bsig_{mm}^{\ast \,s}\right]
\colon \bQs_m \bQs_m \,, \label{e:feapx}
\end{align}
where $\bsig_{mm}^{\ast \,s}$ denotes the matrix product of $s$
copies of $\bsigs_{mm}$. The expansion above is valid provided the
Gaussian averages $\langle \xi_m \rangle$ and $\langle d^s \xi_m/
\,d ({\Qs}_m^{\,2})^s \rangle$ exist for all $s = 1, \ldots,
\infty$. For FENE springs, however,
\begin{gather}
\xi_m = \frac{1}{(1- Q_m^{\ast\,2}/ \bs)} \,,\\
\intertext{and,}
 \frac{d^s \xi_m}{d
({\Qs}_m^{\,2})^s} = \frac{s !}{b^{\ast\,s}} \frac{1}{(1-
Q_m^{\ast\,2}/ \bs)^{s+1}}\,. \label{e:feserterm}
\end{gather}
The Gaussian averages of these functions do not exist since all of
them have non-integrable singularities at $Q_m^{\ast\,2} = \bs$,
whereas the Gaussian distribution has an infinite range. To make
analytical progress, it is necessary to introduce further
approximations. Here, analogous to the Peterlin closure
approximation~\citep{peterlin:jpsb-66},
\begin{gather}
\langle \xi_m  \rangle \text{ is replaced with }
\frac{1}{\langle (1- Q_m^{\ast\,2}/ \bs) \rangle} \,\\
\intertext{and,} \left\langle \frac{d^s \xi_m}{d
({\Qs}_m^{\,2})^s} \right\rangle \text{ is replaced with } \frac{s
!}{b^{\ast\,s}} \frac{1}{\langle (1- Q_m^{\ast\,2}/ \bs)^{s+1}
\rangle}\,.
\end{gather}
Although each of these terms may remain bounded as long as $
\langle Q^{\ast\,2}_m \rangle = \tr \bsigs_{mm} < \bs$, there is a
need to carefully analyze the convergence properties of the
infinite series in Eq.~\eqref{e:feapx} after introducing these
approximations. Here, only the first term in the infinite series
in Eq.~\eqref{e:feapx} is retained to represent spring force
fluctuations. As a result, the following mean-field quadratic
spring potential is used,
\begin{gather}
\overline{S}^\ast_m = \frac{\Hs}{2} \bm{L}_m \colon \bQs_m \bQs_m
\,,
\end{gather}
where,
\begin{widetext}
\begin{align}
 \bm{L}_m &\equiv \frac{1}{\langle(1-  Q_m^{\ast\,2} / \bs )\rangle} \ut
  + \frac{2}{\bs} \frac{1}{\langle (1-  Q_m^{\ast\,2}/
\bs)^2 \rangle} \bsigs_{mm}\, , \notag \\
&= \frac{1}{1-  \tr \bsigs_{mm}  / \bs } \ut +
 \frac{2/\bs}{1 - 2 \tr \bsigs_{mm} / \bs +
[(\tr \bsigs_{mm})^2 + 2 \bsigs_{mm} \colon
\bsigs_{mm}]/b^{\ast\,2}} \bsigs_{mm} \,. \label{e:Ldef}
\end{align}
\end{widetext}
The use of this potential leads to a reduction of the complex
moments in Eq.~\ref{e:xqqfe},
\begin{gather}
\Hs \langle \bQs_i \bQs_m \xi_{m} \rangle = \Hs \bsigs_{im} \cdot
\bm{L}_m \,, \label{e:fecontrib}
\end{gather}
and a mean-field spring force given by,
\begin{gather}
\overline{\bm{F}}^{\textsc{s},\,c}_m = \Hs \bm{L}_m \cdot \bQs_m
\,. \label{e:FENEPG}
\end{gather}

As mentioned above, neglecting the spring force fluctuation term
in $\bm{L}_m$ is equivalent to a consistent-averaging treatment of
the FENE nonlinearity, which leads to the well known FENE-P force
law proposed originally by Bird \emph{et al.\
}\citep{bird:jnnfm-80},
\begin{gather}
\overline{\bm{F}}^{\textsc{s},\,c}_m = \Hs  \frac{1}{1-  \tr
\bsigs_{mm}  / \bs } \bQs_m \, = \Hs \overline{\xi}_m
\bQs_m\label{e:FENEP}
\end{gather}
where, $\overline{\xi}_m = \left[{1-  \tr \bsigs_{mm}  / \bs
}\right]^{-1}$. Incorporation of fluctuations through the Gaussian
approximation leads to the spring force expression in
Eq.~\eqref{e:FENEPG}, denoted here as the ``FENE-PG" spring force
law.

The evolution equation for the second moments in the Gaussian
approximation for free-draining FEBS chains is consequently
\begin{align}
\begin{split}
\bsigs_{ij, \,(1)} = &- \Hs \frac{1}{4} \sum_{m \, =\,1}^{\NS}
\left(\bsigs_{im} \cdot \bm{L}_{m}
 A_{mj} + A_{im} \bm{L}_{m} \cdot \bsigs_{mj} \right) \\ &+
\frac{1}{2} A_{ij} \ut \,, \end{split}\label{e:gaqqfe}
\end{align}
and the polymer stress predicted by the approximation is given by
\begin{gather}
\btaups =  \NS \ut - \Hs \sum_{i = 1}^{\NS} \bsigs_{ii} \cdot
\bm{L}_i \,. \label{e:tauGA}
\end{gather}

It is worth noting that for finite $\bsigs_{mm}$, as $\bs
\rightarrow \infty$, the fluctuation contribution in
Eq.~\eqref{e:Ldef} becomes vanishingly small, $\bm{L}_m
\rightarrow \ut$ and the spring becomes more Hookean-like. On the
other hand, as the bead-spring chain approaches its full
extension, and
\begin{gather}
\bsigs_{mm} \rightarrow \begin{pmatrix}
\bs & 0 & 0 \\ 0 & 0 & 0 \\ 0 & 0 & 0 \\
\end{pmatrix}\,,
\end{gather}
the scalar prefactor in the fluctuation term in Eq.~\eqref{e:Ldef}
approaches $(1/\bs)$, which however becomes negligible compared to
the FENE-P contribution, which diverges. Thus, as chains become
highly stretched, predictions obtained with the FENE-P or FENE-PG
approximations are expected to be nearly identical.

Although the FENE-PG approximation is new and has been developed
in this study, the FENE-P approximation has been studied in great
detail by several workers \citep{wedgewood:iecr-88,
ottinger:jcp-89, wiest:jor-89, wiest:jcp-89, keunings:jnnfm-97,
herrchen:jnnfm-97, brule:jnnfm-93, ghosh:jor-01}. As pointed out
in many of these studies, perhaps the single most important
difference between the exact FENE and the mean-field FENE-P and
FENE-PG approximations lies in the nature of the configurational
probability distribution function. The singularity in the
non-averaged FENE spring force strictly restricts the lengths of
the connector vectors to lie in the domain $[0, Q_0)$. However,
with the Peterlin closure, the solution of the modified
Fokker-Planck equation with FENE-P/PG springs is a Gaussian.
Therefore, there is a non-zero probability that a spring has a
length greater than $Q_0$. This has been clearly illustrated by
Keunings\citep{keunings:jnnfm-97} who used Brownian dynamics
simulations of dumbbells with FENE-P springs to show that in
strong extensional flows, a significant proportion of the
dumbbells had length greater than the ``maximum allowed"
dimensionless extension $Q_0$. However, in all the literature on
the FENE-P model, the mean-squared lengths of the springs are
observed to always be lesser than $Q_0^2$. In other words, the
mean-field FENE-P model appears to satisfy the maximum extension
constraint in an average sense.

It is observed in the studies cited above that predictions with
the FENE-P approximation compare well with the FENE model for
steady-state properties, in both shear and uniaxial extensional
flows. An explanation for the good agreement at steady-state in a
strong flows is based on the observation that under such
conditions most of the springs in the chain are highly stretched.
The exact steady-state distribution function for the length of the
spring in a dumbbell model is sharply peaked around $\sqrt{\langle
Q^2 \rangle}$, and can be well approximated by a $\delta$-function
\citep{lielens:jnnfm-98, wiest:jor-89}. However, during start-up
of shear and extensional flows, predictions with the FENE-P model
for the transient rheological properties deviate strongly from the
exact results of BD simulations prior to the attainment of the
steady-state \citep{herrchen:jnnfm-97, brule:jnnfm-93}.

There have been several previous attempts to improve the FENE-P
approximation. In conditions where springs can be highly
stretched, the true spring length distribution is better
approximated as the sum of a uniform distribution for smaller
values of $Q$ and a $\delta$-function located at a large $Q$
closer to (but less than) $Q_0$. This FENE-L approximation, and
its three-dimensional counterpart, are shown to be more accurate
than the FENE-P model, particularly in predicting the phenomenon
of stress-conformational hysteresis \citep{lielens:jnnfm-98,
lielens:jnnfm-99, sizaire:jnnfm-99}. Grban \emph{et al.\
}\citep{gorban:jnnfm-01} show that an entropic argument can be
used to improve upon the FENE-P approximation by first expanding
the exact FENE potential in a Taylor's series expansion about
$\langle Q^2 \rangle$ and then systematically regrouping terms in
the series before truncating it. This approach gives rise to a
hierarchy of spring potentials of increasing complexity, the
simplest of which is shown to be the quadratic potential
corresponding to the FENE-P force law. By using the next higher
order potential (which is denoted as the ``FENE-P+1" potential by
Gorban \emph{et al.\ }) in the hierarchy, these authors
demonstrate with a one-dimensional dumbbell model that accuracy is
significantly improved in an unsteady elongational flow. In
Section~\ref{s:randd} below, the accuracy of the FENE-PG model
proposed in this study is evaluated by comparison with exact
Brownian dynamics simulations.

The choice of parameter values $\Hs$ and $\bs$ for the approximate
FENE-P and FENE-PG models is made here consistently with the
procedure used in the case of the FENE force law; namely, the
equilibrium mean square end-to-end distance of a spring predicted
by the model is equated to $\bK^2 \NKS$, while the fully stretched
length is equated to $\bK \NKS$. Further, as before, $\elS^2 =
R_{\textsc{s}}^2/3 = \bK^2 \NKS/3$, and the corresponding time
scale $\lS$, are used as the basic scales for obtaining equations
and properties in dimensionless form. These assumptions lead to
$\bs = Q_0^2/\elS^2 = 3 \NKS$, as in the earlier case. From
Eq.~\eqref{e:gaqqfe}, it is clear that the equilibrium second
moments $\bsigs_{ij,\,\textrm{eq}}$ predicted by the Gaussian
approximation for free-draining FEBS chains, are solutions to the
following set of algebraic equations:
\begin{align}
\begin{split}
\bm{0} = &- \frac{\Hs}{4} \sum_{m = 1}^{\NS} \left(
\bsigs_{im,\,\textrm{eq}} \cdot \bm{L}_{m,\,\textrm{eq}} A_{mj} +
A_{im} \bm{L}_{m,\,\textrm{eq}} \cdot \bsigs_{mj,\,\textrm{eq}}
\right) \\
&+ \frac{1}{2} A_{ij} \ut \, .
\end{split}
\end{align}
At equilibrium, since $\bsigs_{ij,\,\textrm{eq}} =
\sigma^\ast_{ij,\,\textrm{eq}} \ut$ and $\bm{L}_{m,\,\textrm{eq}}
= L_{m,\,\textrm{eq}} \ut$, the equation above can be rearranged
into the following set of equations,
\begin{gather}
0 = \sum_{m = 1}^{\NS} \left( U_{im} A_{mj} + A_{im} U_{mj}
\right) \label{e:fdeqbm1}
\end{gather}
where,
\begin{gather}
U_{ij} \equiv \Hs \sigma^\ast_{ij,\,\textrm{eq}}
L_{j,\,\textrm{eq}} - \delta_{ij} \label{e:fdeqbm2}
\end{gather}
Since the Rouse matrix is non-singular, the solution to
Eq.~\eqref{e:fdeqbm1} is obtained by setting $U_{ij}=0$. This
implies, $\bsigs_{ij,\,\textrm{eq}} = \delta_{ij} \ut$, and $\Hs
\bm{L}_{j,\,\textrm{eq}} = \ut$. With all springs being identical
at equilibrium , $\bm{L}_{j,\,\textrm{eq}} = L_\textrm{eq} \ut$,
and therefore
\begin{gather}
\Hs L_\textrm{eq} = 1 \,. \label{e:HsLeq}
\end{gather}
Using $\tr \bsigs_{jj\,\textrm{eq}} = 3$, and $\bs = 3 \NKS$, it
can be shown using Eq.~\eqref{e:Ldef} that for FENE-PG springs,
\begin{gather}
\Hs = \frac{3 \NKS^3 - 9 \NKS^2 + 13 \NKS - 5}{3 \NKS^3 - 4 \NKS^2
+ 5 \NKS} \, .
\end{gather}
For FENE-P springs, $L_\textrm{eq} =
\overline{\xi}_{m,\,\textrm{eq}}$ (for all $m$), where
$\overline{\xi}_m$ was introduced in Eq.~\eqref{e:FENEP} earlier.
As a result,
\begin{gather}
\Hs =  \frac{1}{\overline{\xi}_{m,\,\textrm{eq}}} = 1 -
\frac{1}{\NKS}\,. \label{e:HsFENEP}
\end{gather}

\subsection{\label{s:gafebs} FEBS chains with HI}

The dimensionless evolution equation for the second-moments of a
bead-spring chain model with springs obeying a mean-field
potential given by Eq.~\eqref{e:mfsprpot}, and with fluctuating HI
between the beads, is given by,
\begin{align}
\begin{split}
\QQs{i}{j}_{(1)} = &- \frac{\Hs}{4} \sum_{m=1}^{\NS} \langle
\bQs_i \bQs_m \cdot \bm{L}_m \cdot \bA_{mj} \\ & + \bA_{im} \cdot
\bm{L}_m \cdot \bQs_m \bQs_j \rangle + \frac{1}{2} \langle
\bA_{ij} \rangle\,, \end{split} \label{e:qqmfsprhi}
\end{align}

Since the tensors $\bm{L}_m$ are not direct functions of the
connector vectors, expectations of the form $\langle \bQs_i \bQs_m
\cdot \bm{L}_m \cdot (\zeta \bO_{mj}) \rangle$ appearing in the
equation can be simplified by assuming that the configurational
distribution function is Gaussian and using Wick's decomposition
rule as in the case of free-draining FEBS chains above. The final
equation thus obtained for the approximate second-moments
$\bsigs_{ij}$ is,
\begin{align}
\begin{split} \bsigs_{ij,\,(1)} = &- \frac{\Hs}{4} \sum_{m=1}^{\NS} \left[
\bsigs_{im} \cdot \left(\bm{L}_m \cdot \bAb_{mj} +\bm{\Delta}_{mj}
\right) \right.
\\ &+  \left.  \left( \bAb_{im}
\cdot \bm{L}_m + \bm{\Delta}_{mi}^\textsc{t}  \right) \cdot
\bsigs_{mj}\right] + \frac{1}{2} \bAb_{ij} \,,
\end{split}
\label{e:gapgqq}
\end{align}
where, the tensor $\bAb_{ij}$ describes contributions arising from
averaging the Oseen-Burgers tensor with the non-equilibrium
distribution function, while the tensor $\bm{\Delta}_{ij}$
represents contributions due to fluctuations in HI. Since the
expressions for these tensors are lengthy and not very
illuminating in themselves, they are given by Eqs.~\eqref{e:babij}
and~\eqref{e:deltaij} in Appendix~\ref{s:appA}. The important
point to note here is that $\bAb_{ij}$ and $\bm{\Delta}_{ij}$ are
completely expressible in terms of the second moments of the
Gaussian distribution, and Eq.~\eqref{e:gapgqq} is closed with
respect to the tensors $\bsigs_{ij}$.

At equilibrium, all the tensorial quantities in the equations
above are isotropic. Furthermore, the HI fluctuation tensors
$\bm{\Delta}_{ij}$ vanish  under isotropic conditions (see
Appendix~\ref{s:appA}). Consequently, one obtains an equation of
the same form as Eq.~\eqref{e:fdeqbm1}:
\begin{gather}
0 = \sum_{m=1}^{\NS} [U_{im} \widetilde{A}_{mj} +
\widetilde{A}_{im} U_{mj}] \,, \label{e:gaqqeq}
\end{gather}
in which the values of $U_{ij}$ are the same as in
Eq.~\eqref{e:fdeqbm2} for the \emph{free-draining} model with
FENE-PG springs, and $\widetilde{A}_{im}$ are elements of the
modified Rouse matrix (Eq.~\eqref{e:zbabs}),
\begin{widetext}
\begin{gather}
\langle \bA_{ij} \rangle_\textrm{eq} = \widetilde{A}_{ij} \ut =
\left[ A_{ij}  + \sqrt{2} h^\ast \,
 \left( \frac{2}{\sqrt{|i - j|}} - \frac{1}{\sqrt{|i - j + 1|}} -
\frac{1}{\sqrt{|i - j - 1|}} \right) \right] \ut, \label{e:zbabs}
\end{gather}
\end{widetext} where, the convention $1/\sqrt{0} = 0$ has been
used. Since the modified Rouse matrix is also non-singular, the
equilibrium solution is once again obtained by setting $U_{ij} =
0$. Therefore, the equilibrium solution obtained for the Gaussian
approximation for FENE-PG chains with HI is identical to that
obtained for free-draining FENE-PG chains. In other words, the
heuristic two-step procedure outlined above to achieve closure in
the model combining both FE and HI ensures that the equilibrium
solution for the second-moments is independent of HI.

Nearly all the closure approximations introduced to date for the
second-moments $\QQs{i}{j}$ can be derived by dropping appropriate
terms in Eq.~\eqref{e:gapgqq}. For instance, the Zimm model is
obtained by (i) using a Hookean spring force law, $\bm{L}_m =
\ut$, (ii) replacing the non-equilibrium averaged Oseen tensor
contribution $\bAb_{ij}$ with the modified Rouse matrix
$\widetilde{A}_{ij}$ (which arises due to equilibrium averaging
the Oseen tensor), and (iii) setting $\bm{\Delta}_{ij} = \bm{0}$.
Ignoring fluctuations in hydrodynamic interactions in a model with
Hookean springs ($\bm{\Delta}_{ij} = \bm{0}$ and $\bm{L}_m = \ut$)
leads to the consistent averaging (CA) approximation
\citep{ottinger:jcp-87c}, while retaining fluctuations in HI leads
to the Gaussian approximation (GA) \citep{ottinger:jcp-89}.
Setting $\bm{\Delta}_{ij} = \bm{0}$ and $\bm{L}_m =
\overline{\xi}_m \ut$ simultaneously (which implies fluctuations
in both HI and the spring force are ignored), leads to a model
combining the Consistent-Averaging treatment of HI with the FENE-P
approximation for the spring force. The predictions of this CA-P
model for properties in shear flows have been studied in detail by
Wedgewood and \"{O}ttinger\citep{wedgewood:jnnfm-88}. On the other
hand, using only one of the two conditions leads to a model in
which fluctuations in the corresponding phenomenon alone are
ignored, leading to the CA-PG model or the GA-P model. The model
incorporating fluctuations in both nonlinear phenomena is referred
to here as the GA-PG model. Note that in all these models, the
polymer stress is given by Eq.~\eqref{e:tauGA}, with the
appropriate choice of $\bm{L}_i$.

In discussing predictions obtained using the various approximate
models, it is convenient to use acronyms in which the first two
letters refer to the approximation used for HI, and the last
letters refer to that used for FE. Table~\ref{t:apxcode} lists the
letter codes used for easy reference. Note that in this scheme,
the Rouse model is denoted as the FD-H model, while the Zimm model
is denoted as EA-H.

\begin{table}[!t]
\caption{\label{t:apxcode} Acronyms used for referring to
approximate models.} \addtocounter{table}{-1}
\begin{longtable}[c]{p{2.5in} c}
&  \kill \hline \hline
& Acronym \\
\hline
\multicolumn{1}{c}{Treatment of HI} \\[1mm]
Free-draining & FD  \\
Equilibrium-averaging & EA \\
Consistent-averaging & CA \\
Gaussian approximation & GA \\
\multicolumn{1}{c}{Treatment of FE} \\[1mm]
Hookean springs & H \\
FENE-P & P \\
FENE-PG & PG \\[6pt]
\hline
\end{longtable}
\end{table}

\subsection{\label{s:diag} Diagonalization}

Previous studies have shown that considerable savings in
computational time, without a significant loss in accuracy, may be
achieved by introducing a \textit{diagonalization approximation}
\citep{kisbaugh:jnnfm-90, prakash:jnnfm-97}. The most recent of
these approximations, carried out in the context of the Gaussian
approximation for Rouse chains with HI, is the ``Two-Fold Normal"
(TFN) approximation of Prakash and
\"{O}ttinger\citep{prakash:jnnfm-97} (named as such because of two
different assumptions of ``normality"--a Normal distribution, and
a transformation to normal coordinates). Prakash and \"{O}ttinger
showed that for Rouse chains with HI, the predictions of the TFN
approximations for linear viscoelastic properties and steady-state
material functions in shear are close to those obtained with the
original Gaussian approximation. Here, we examine the
effectiveness of making a similar approximation for FENE-P chains
with HI.

The first step in introducing a diagonalization approximation is
mapping the set of the physical connector vectors $\{\bQs_i | i =
1,\dots, \NS\}$ onto a set of normal coordinates $\{\bQ_p' | p =
1,\dots, \NS\}$ using the following transformation,
\begin{gather}
\bQ_p' = \sum_{i = 1}^{\NS} \Pi_{ip}^\textsc{z} \bQs_i\,,
\label{e:zortho}
\end{gather}
where $\Pi_{ip}^\textsc{z}$ are the elements of the orthogonal
matrix whose columns are formed by the eigenvectors of the
modified Rouse matrix,
\begin{gather}
\sum_{i,j = 1}^{\NS} \Pi_{ip}^\textsc{z} \widetilde{A}_{ij}
\Pi_{jq}^\textsc{z} = \delta_{pq}\, \widetilde{a}_p\,,
\end{gather}
in which $\widetilde{a}_p$ is an eigenvalue of the modified Rouse
matrix. The orthogonal matrix corresponding to the Rouse matrix is
used for free-draining models without HI, for effecting the
transformation to normal coordinates. Using the orthogonality
relation
\begin{gather}
\sum_{k = 1}^{\NS} \Pi_{mk}^\textsc{z} \Pi_{nk}^\textsc{z} =
\delta_{mn} = \sum_{k = 1}^{\NS} \Pi_{km}^\textsc{z}
\Pi_{kn}^\textsc{z}\,,
\end{gather}
the evolution equations for the covariances of the normal
coordinates $\bsig_{pq}'$ defined as,
\begin{gather}
\bsig_{pq}' = \sum^{\NS}_{i,j = 1} \Pi_{ip}^\textsc{z} \bsigs_{ij}
\Pi_{jq}^\textsc{z} \,, \label{e:nmcov}
\end{gather}
can be shown to be,
\begin{align}
\begin{split}
\bsig_{pq,\,(1)}'  = &- \frac{1}{4} \sum_{u = 1}^{\NS} \left[
\bsig_{pu}'\cdot (\bm{X}_{uq} + \bm{Y}_{uq}) + (\bm{X}_{up} +
\bm{Y}^\textsc{t}_{up}) \cdot\bsig_{uq}' \right] \\&+ \half
\bm{Z}_{pq} \,,
\end{split}\label{e:tfnpqqfull}
\end{align}
where the auxiliary tensors are defined by,
\begin{gather}
\bm{Z}_{pq} \equiv \sum_{i,j = 1}^{\NS} \Pi_{ip}^\textsc{z}
\bAb_{ij}
\Pi_{jq}^\textsc{z} \,,\label{e:Ztendef}\\
\bm{X}_{pq} \equiv \Hs \,\sum_{i,j = 1}^{\NS} \Pi_{ip}^\textsc{z}
  \overline{\xi}_i \bAb_{ij} \Pi_{jq}^\textsc{z} \,,
  \label{e:Xtendef}\\
\bm{Y}_{pq} \equiv \Hs \,\sum_{i,j = 1}^{\NS} \Pi_{ip}^\textsc{z}
\overline{\xi}_i \bm{\Delta}_{ij} \Pi_{jq}^\textsc{z} \,.
\label{e:Ytendef}
\end{gather}

Following \citeauthor{prakash:jnnfm-97}, we \emph{assume} that the
normal coordinates covariance matrix is diagonal,
\begin{gather}
\bsig_{pq}' \approx \delta_{pq} \bsig_{p}' \,, \label{e:tfnpdiag}
\end{gather}
This leads to the following set of equations for the normal
coordinate variances,
\begin{gather}
\bsig_{p,\,(1)}'  = - \frac{1}{4} \left[ \bsig_{p}'\cdot
(\bm{X}_{p} + \bm{Y}_{p}) + (\bm{X}_{p} + \bm{Y}^\textsc{t}_{p})
\cdot\bsig_{q}' \right] + \half \bm{Z}_{p} \,, \label{e:tfnpqq}
\end{gather}
where the notation $(\dots)_{p} = (\ldots)_{pp}$ is used. While
\citeauthor{prakash:jnnfm-97} argue that it is not necessary to
make the additional diagonalization assumptions,
\begin{gather}
\bm{Z}_{pq} \approx \bm{Z}_p \delta_{pq}; \quad  \bm{X}_{pq}
\approx \bm{X}_p \delta_{pq} ; \quad \bm{Y}_{pq} \approx \bm{Y}_p
\delta_{pq} \label{e:dcapcons}
\end{gather}
in order to derive Eq.~\eqref{e:tfnpqq}, Eq.~\eqref{e:tfnpqqfull}
shows that for the time derivatives of the off-diagonal components
of the normal-mode covariances to remain negligible far from
equilibrium, it is necessary for both Eq.~\eqref{e:tfnpdiag} and
Eqs.~\eqref{e:dcapcons} to be true.

Equation~\eqref{e:tfnpqq} represents a set of $\NS$ ODE's that are
coupled and nonlinear. The coupling comes about because the
auxiliary tensors $\bm{Z}_p$, $\bm{X}_p$ and $\bm{Y}_{p}$ depend
on all the variances $\bsigs_{ij}$. For instance, the FENE-P
factors appearing in these terms are evaluated as
\begin{gather}
\overline{\xi}_p = \frac{1}{1-   \left[\displaystyle{\sum_{p =
1}^{\NS} \Pi_{pm}^{\textsc{z}\,2}}\, \tr (\bsig'_{m} )\right] /
\bs } \,.
\end{gather}
As a result of the coupling, the CPU-time per time step scales as
$\NS^3$, even though the number of equations is only $\NS$. For
every $p$ in Eq.~\eqref{e:tfnpqq}, each calculation of $\bm{Z}_p$,
$\bm{X}_p$ and $\bm{Y}_{p}$ individually involves a double
summation over all $i$ and $j$ in Eqs.~\eqref{e:Ztendef},
\eqref{e:Xtendef} and \eqref{e:Ytendef}. In the calculation of the
variances $\bm{S}_{\mu \nu}^\ast$ using Eq.~\eqref{e:Sneq}, each
of the $\bsigs_{ij}$ requires a summation over all $k$, as can be
seen from,
\begin{gather}
\bsigs_{ij} = \sum_{k = 1}^{\NS} \Pi_{ik}^\textsc{z} \bsig_{k}'
\Pi_{jk}^\textsc{z} \,. \label{e:znm2phy}
\end{gather}
Since the values of $\bsigs_{ij}$ can be stored at the beginning
of each time-step, the total cost of generating the super-matrix
containing the $\bsigs_{ij}$ tensors scales as $\NS^3$. The cost
of the calculating $\bm{S}_{\mu \nu}^\ast$ using
\begin{gather}
\bm{S}_{\mu \nu}^\ast = \bm{S}^\ast_{\mu, \nu-1} + \sum_{i =
\mu}^{\nu-2} (\bsigs_{i,\nu-1} + \bsigs_{\nu-1,i}) +
\bsigs_{\nu-1,\nu-1}, \quad \mu < \nu,
\end{gather}
can also be shown to scale as $\NS^3$. Thus, as far as the
CPU-time scaling is concerned, the diagonalization assumption does
not offer an advantage over the original approximations in the
calculation of properties in unsteady flows using bead-spring
models with HI. However, the reduction in the number of equations
to be integrated greatly reduces the prefactor in the CPU-time's
power-law dependence on $\NS$, and as will be shown in
section~\ref{s:tfn}, leads to a significant reduction in CPU time.

The diagonalized versions of the models will be referred to with a
``D" prefixed to the acronym representing the original
approximation, \emph{e.g.} DFD-P, DCA-P \emph{etc.} The exception
to this rule is the TFN-P model, combining the TFN approximation
for HI with FENE-P springs.

Before presenting results, the next two sections briefly introduce
the main material functions examined here, and the numerical
schemes used to obtain them.

\section{\label{s:mf} Material functions}

Attention is restricted here to some instances of simple shear and
uniaxial extensional flows. Expressions for the material functions
in these flows, in terms of components of the stress tensor, are
discussed below.

\subsection{\label{s:sf} Shear flow}
For a simple homogeneous shear flow characterized by the shear
rate $\dot{\gamma}$, the continuum solvent's velocity components
in Cartesian coordinates are given by $v_x = \gdot(t) \,y$, $v_y =
0$ and $v_z = 0$. The tensor $\bks = \lS \bm{\nabla} \bv$ in
Cartesian coordinates is
\begin{gather}
\bks(\ts)=  \gdots (\ts)  \begin{pmatrix}
0 & 1 & 0 \\ 0 & 0 & 0 \\ 0 & 0 & 0 \\
\end{pmatrix} \, , \label{e:shrka}
\end{gather}
where $\gdots (\ts)= \gdot (t) \lS$ is the time dependent
dimensionless shear rate. Here, the time dependence of solution
properties is examined for the sudden imposition of steady shear
flow:
\begin{align}
\gdots(\ts) =
\begin{cases}
0\,, &\quad \ts < 0\,, \\
\gdots &\quad \ts \geq 0\,.
\end{cases}
\label{e:flo1}
\end{align}
While the imposed strain rate $\gdots$ at $\ts \geq 0$ is a
constant, the solution as a whole is in an unsteady state since
the properties of the solution change in response to the
imposition of the flow. Steady-state is reached when all
properties reach time-independent values which are functions
solely of the imposed strain rate and the other physical
parameters characterizing the solution.

For a simple shear flow defined in Eq.~\eqref{e:flo1}, the
rheological properties of interest are the polymer's contribution
to the dimensionless viscosity $\etaps$, the dimensionless first
normal-stress difference coefficient $\PsiIs$, and the
dimensionless second normal-stress difference coefficient
$\PsiZs$, which are respectively defined by the following
relations:
\begin{align}
\etaps\, & \equiv\, -\,
\frac{\tau^\ast_{\text{p},\,yx}}{\gdots}, \label{e:seta}\\
\PsiIs\, & \equiv\, -\, \frac{\tau^\ast_{\text{p},\,xx}-
\tau^\ast_{\text{p},\,yy}}{\dot{\gamma}^{\ast\,2}}, \label{e:psi1}\\
\PsiZs\, & \equiv\, -\, \frac{\tau^\ast_{\text{p},\,yy}-
\tau^\ast_{\text{p},\,zz}}{\dot{\gamma}^{\ast\,2}}\,.
\label{e:psi2}
\end{align}

\subsection{\label{s:ef} Extensional flow}
In a uniaxial extensional flow with an extension-rate $\edot$, the
velocity field is specified by $v_x = \edot(t) x$, $v_y =
-(\edot(t)/2) y$ and $v_z = -(\edot(t)/2) z$. Therefore,
\begin{gather}
\bks(\ts) = \edots(\ts) \begin{pmatrix}
  1 & 0 & 0 \\ 0 & -\frac{1}{2} & 0 \\ 0 & 0 & -\frac{1}{2} \\
\end{pmatrix} \, , \label{e:uaeka}
\end{gather}
where $\edots (\ts) = \edot (t) \lS$ is the time dependent
dimensionless extension-rate. Here, the time dependence of
solution properties is examined for the following kinds of
unsteady extensional flows.
\begin{itemize}
\item[]Sudden imposition of steady shear or uniaxial extensional
flow:
\begin{align}
\edots(\ts) =
\begin{cases}
0\,, &\quad \ts < 0\,, \\
\edots \,, &\quad \ts \geq 0\,.
\end{cases}
\label{e:flo3}
\end{align}
\item[]Sudden imposition followed by cessation of steady uniaxial
extensional flow:
\begin{align}
\edots(\ts) =
\begin{cases}
0\,, &\quad \ts < 0\,, \\
\edots \,, &\quad 0 \leq \ts \leq \ts_\textrm{max} \text{ (the stress-growth phase)}\,, \\
0\,, &\quad \ts > \ts_\textrm{max} \text{ (the stress-relaxation
phase)}\,.
\end{cases}
\label{e:flo2}
\end{align}
\end{itemize}
The time-dependence of properties in an unsteady state is
expressed in terms of the strain measure $\epsilon = \edots \ts$.

In the case of a uniaxial extensional flow specified by
Eq.~\eqref{e:flo3}, the polymer contribution to the dimensionless
extensional viscosity, denoted as $\eetaps$, is given by
\begin{align}
\eetaps\, & \equiv\, -\, \frac{\tau^\ast_{\text{p},\,xx}-
\tau^\ast_{\text{p},\,yy}}{\edots}\,. \label{e:eeta}
\end{align}
For the flow pattern described in Eq.~\eqref{e:flo2}, rheological
data is presented in this work in terms of $N_1^\ast$, the
magnitude of the polymer's contribution to the dimensionless first
normal stress difference:
\begin{align}
N_{1,\textrm{p}}^\ast = |\tau_{\text{p},\,xx}^\ast -
\tau_{\text{p},\,yy}^\ast|\,. \label{e:N1}
\end{align}

We note here that the characterization of a shear rate as ``low"
or ``high", depends on whether the Weissenberg number is much
lower, or greater than unity. The Weissenberg number is defined as
the product of the shear (or extension) rate and a time-scale
characterizing the large scale dynamics of isolated chains near
equilibrium. As is well known, this characteristic time-scale is
proportional to $N^2$ for free-draining chains, whereas in chains
with HI, it is proportional to $N^{3/2}$. The Weissenberg number
in these two cases thus scales as $\gdots N^2$ and $\gdots
N^{3/2}$, respectively. For $N = 20$, a shear rate of $\gdots \sim
1$ therefore represents a large strain rate, as will become
evident in the results below. In extensional flows, the critical
strain rate for the coil-to-stretch transition (discussed in
section \ref{s:ref}) marks the boundary between small and large
extension rates.

\section{\label{s:nm} Numerical method}

In the case of the un-diagonalized approximations, the total
number of second moments, and therefore the number of equations,
is equal to $9 \NS^2$. However, $\bsigs_{ij} =
\bsig_{ji}^{\ast\,\textsc{t}}$. Further, the second moments must
not change if the bead numbering is reversed. Thus, $\bsigs_{ij} =
\bsigs_{\NS - i,\NS - j}$. Therefore, the number of independent
second moments reduces by a factor of nearly 1/4
\citep{ottinger:jcp-89}. In the diagonalized approxaimations, on
the other hand, the number of equations for the components of the
normal mode variances is just $9 \NS$

The symmetry of the imposed flow field can be used to reduce the
number of components for all the tensorial quantities in the
equations. In shear flows, the second moment tensors take the
following form:
\begin{align}
\bsig = \begin{pmatrix} \sigma_1 & \sigma_4 & 0 \\
\sigma_5 & \sigma_2 & 0 \\
0 & 0 & \sigma_3 \end{pmatrix} \,.
\end{align}
In uniaxial extensional flows, in both stress growth and
relaxation phases, all tensorial quantities in the approximations
have the form,
\begin{align}
\bsig = \begin{pmatrix} \sigma_1 & 0 & 0 \\
0 & \sigma_2 & 0 \\
0 & 0 & \sigma_2 \end{pmatrix} \,.
\end{align}

The exploitation of the symmetries above to reduce the computation
of the fourth rank tensorial function $\bm{K}(\mathbf{s})$ has
been discussed thoroughly by Zylka\citep{zylka:jcp-91}. In
general, for any symmetric tensor $\mathbf{s}$, only 21 of the 81
components of the $\bm{K}$ function are non-zero in a transformed
coordinate system $C$ in which the tensor $\mathbf{s}$ has a
diagonal representation. Out of these 21 components, only 9
require independent numerical evaluation. The components
$K_{1111}$, $K_{2222}$ and $K_{3333}$ can be used to calculate the
remaining 12 non-zero components. Furthermore, the function
$\bm{K}$ always appears in the equations as $\bm{K} \colon \bsig$.
In shear flows, this means that only 14 components of the $\bm{K}$
function in the laboratory fixed coordinate system are required to
be evaluated. In uniaxial extensional flows, it can be shown that
only $K_{xxxx}$ (where $x$ is the direction of stretching) is
required. As shown by Zylka\citep{zylka:jcp-91}, components of the
$\bm{H}(\mathbf{s})$ and $\bm{K}(\mathbf{s})$ functions in the
coordinate system $C$ are efficiently evaluated using elliptic
integrals.

The initial condition for the time integration of the ODE's
starting from equilibrium is $\bsigs_{ij,\,\textrm{eq}} =
\delta_{ij} \ut$ for the un-diagonalized approximations.
Similarly, in the case of the diagonalized approximations,
\begin{gather}
\bsig_{p,\,\textrm{eq}}' = \sum^{\NS}_{i,j = 1}
\Pi_{ip}^\textsc{z} \bsigs_{ij,\,\textrm{eq}} \Pi_{jp}^\textsc{z}
 = \sum^{\NS}_{i,j = 1} \Pi_{ip}^\textsc{z} (\delta_{ij}\ut)
\Pi_{jp}^\textsc{z} = \ut \,.
\end{gather}
As pointed out in the preceding sections, at equilibrium
$\bAb_{ij,\,\textrm{eq}} = \widetilde{A}_{ij} \ut$,
$\Delta_{ij,\,\textrm{eq}}  = \bm{0}$, $\Hs \bm{L}_{i,
\textrm{eq}} = \ut$.

Since predictions are obtained only for shear and uniaxial
extensional flows in this study, separate computer codes were
written for the two different flows, to take advantage of the
reduction in the components due to their individual flow
symmetries. A variable-step Adams method (the D02CJF routine in
the NAG library) is used for the integration of the ODE's.

Steady-state solutions were obtained by setting the time
derivatives in the ODE's to zero, and then using a Wegstein
successive substitution algorithm to solve the set of nonlinear
equations. Solutions are obtained for several values of the (shear
or extensional) strain rate. The converged solution for the set of
$\bsigs_{ij}$ (or $\bsig'_{p}$) at any strain rate is used as the
initial guess for the next higher (or lower) strain rate for which
a solution is desired. It is found that large steps in the strain
rates could lead to unphysical, non-positive-definite solutions
for the second moments\citep{wedgewood:jnnfm-88,
kisbaugh:jnnfm-90}. Such behaviour is avoided by manually
controlling the size of the steps in strain rate.

The Brownian dynamics simulations of Rouse and FENE chains with
(and without) RPY-HI performed in this study use a modification of
the semi-implicit predictor-corrector scheme outlined recently by
several workers \citep{somasi:jnnfm-02, hsieh:jnnfm-03,
prabhakar:jnnfm-04}, where the error between the predictor and
corrector steps is used to adaptively control the time-step size
in every simulation.

\section{\label{s:randd} Results and Discussion}

Results for shear flows and extensional flows are discussed
separately in the respective sections below. In each case, results
are presented for (i) models with non-linear force laws, but with
no HI, (ii) models with HI incorporated, but with a linear Hookean
force law, and (iii) models with non-linear force laws and HI
incorporated. In particular, in case (i), results obtained with
the new FENE-PG force law introduced here are compared with
results obtained with the FENE-P force law and with BDS, in case
(ii) (which involves only models that have been introduced
previously), results for certain material functions that have not
been predicted earlier, are presented, and in case (iii), the
combined influence of both phenomena are examined, aided by the
results obtained in cases (i) and (ii).

\subsection{\label{s:rsf} Shear flow}

\subsubsection{\label{s:rsfa} Startup of shear flow}

\begin{figure}[t]
\centerline {\resizebox{7cm}{!}{\includegraphics{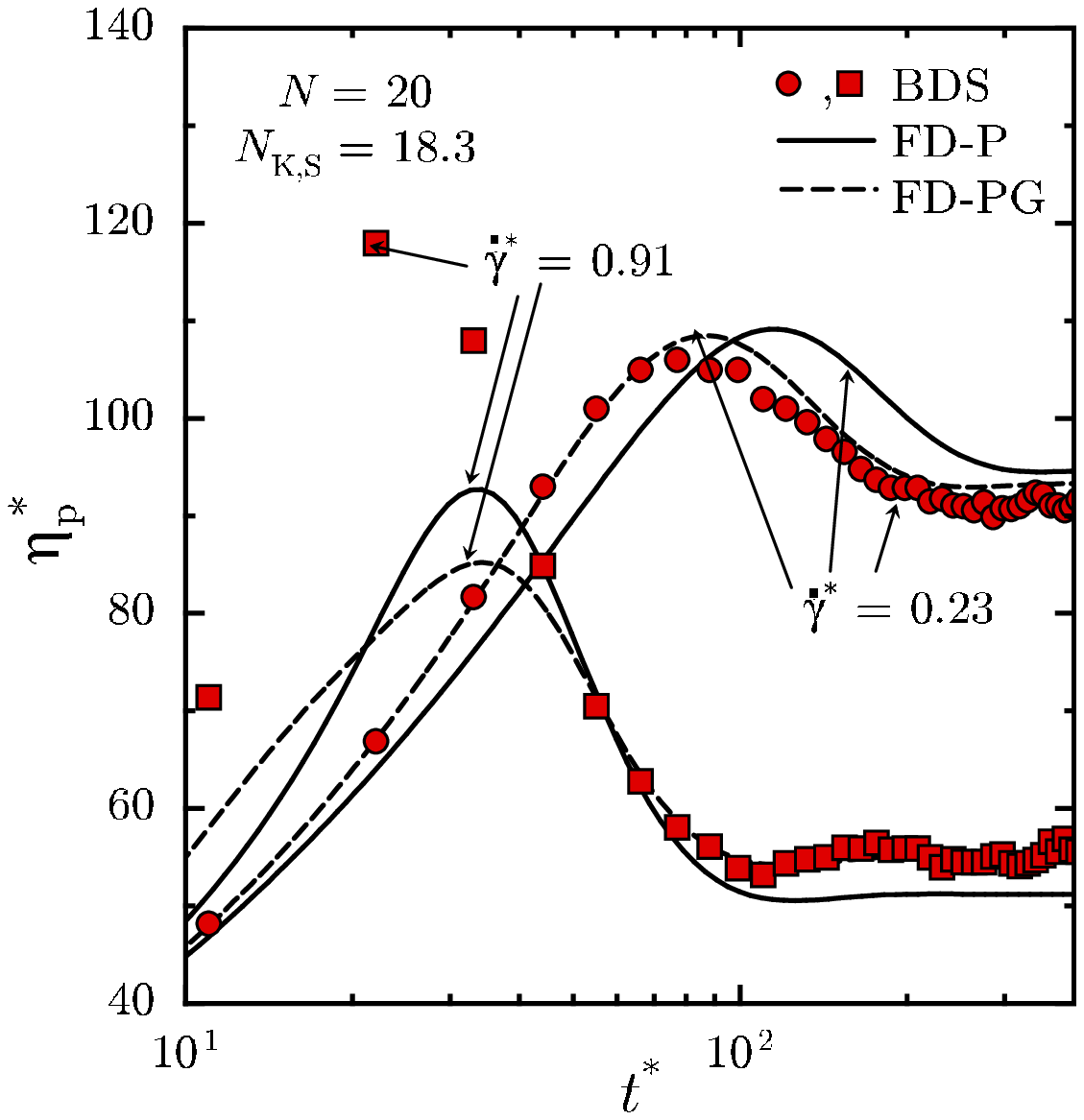}}}
\centerline {\resizebox{7cm}{!}{\includegraphics{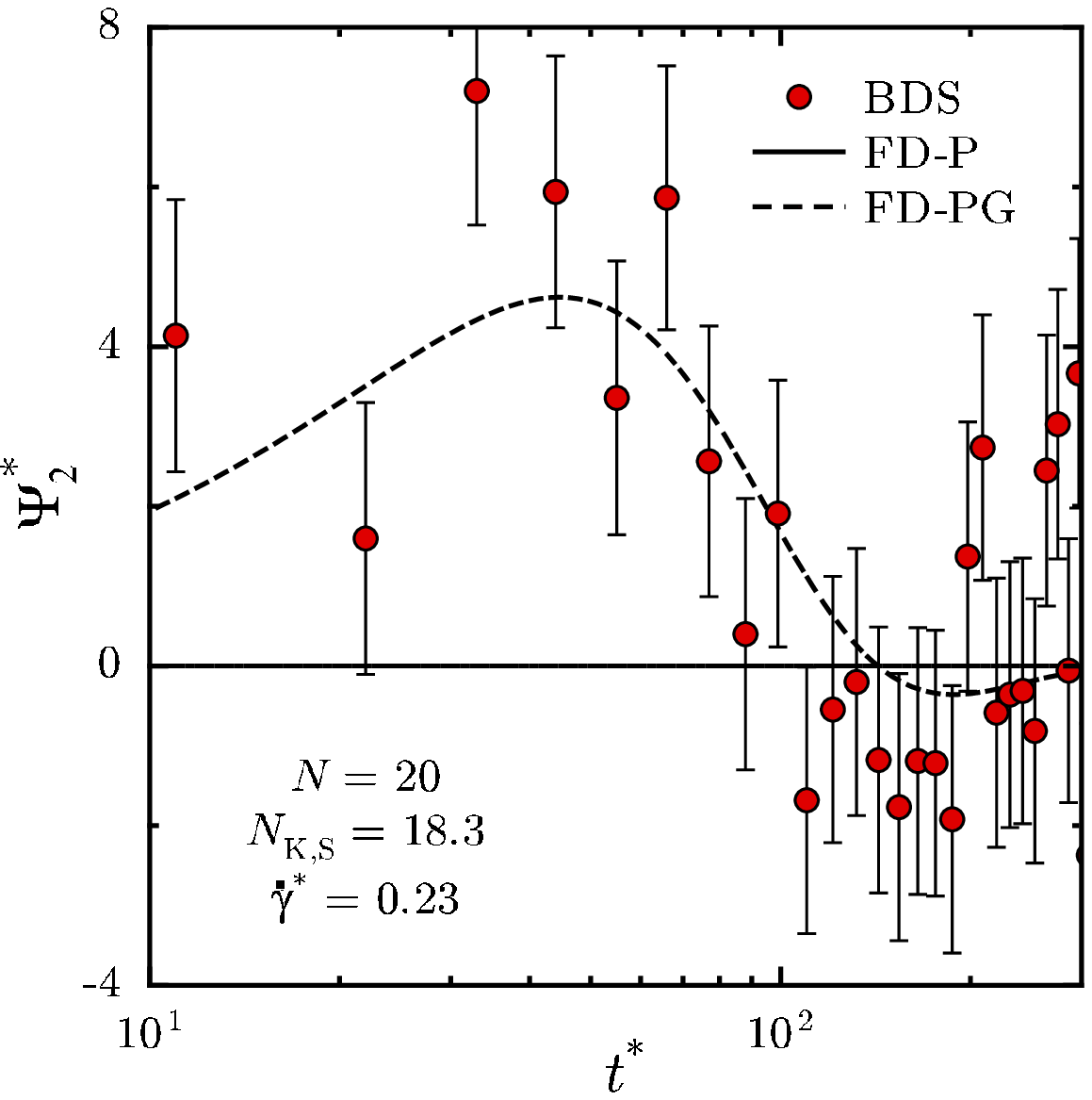}}}
\caption{\label{f:fdusetapsi2} Growth of (a) viscosity and (b)
second normal-stress difference coefficient, for free-draining
FEBS chains during start-up of steady shear flow.}
\end{figure}

The performance of the FENE-P and FENE-PG approximations for
free-draining FEBS chains in the prediction of unsteady behaviour
in shear flows is first considered. At low shear rates close to
equilibrium, finite chain extensibility has little influence on
the dynamic behaviour predicted by the models. Therefore, the
predictions of the free-draining FEBS models are close to those
predicted with the Rouse model. At higher shear rates, earlier
studies with free-draining FENE-P dumbbells
\citep{herrchen:jnnfm-97} and chains with $N = 10$
\citep{brule:jnnfm-93} show that this approximation leads to
predictions for unsteady shear properties that agree only
qualitatively with the exact results obtained using FENE springs.
Figures~\ref{f:fdusetapsi2}~(a) and~(b) show that this trend is
continued for 20-bead chains. The incorporation of fluctuations in
the spring force with the use of the FENE-PG approximation leads
to a clear improvement in the quality of the predictions for all
properties at moderate shear rates (represented by the curves for
$\gdots = 0.23$). As the dimensionless shear rate approaches
unity, however, the FENE-PG approximation also begins to deviate
from the exact BDS results. Interestingly, it is seen that the
FENE-PG force law leads to the prediction of a transient non-zero
positive second-normal stress difference coefficient which
approaches zero at steady-state, in good agreement (within errors
in the simulations' results) with the results of simulations.

\begin{figure}[t]
\centerline {\resizebox{7cm}{!}{\includegraphics{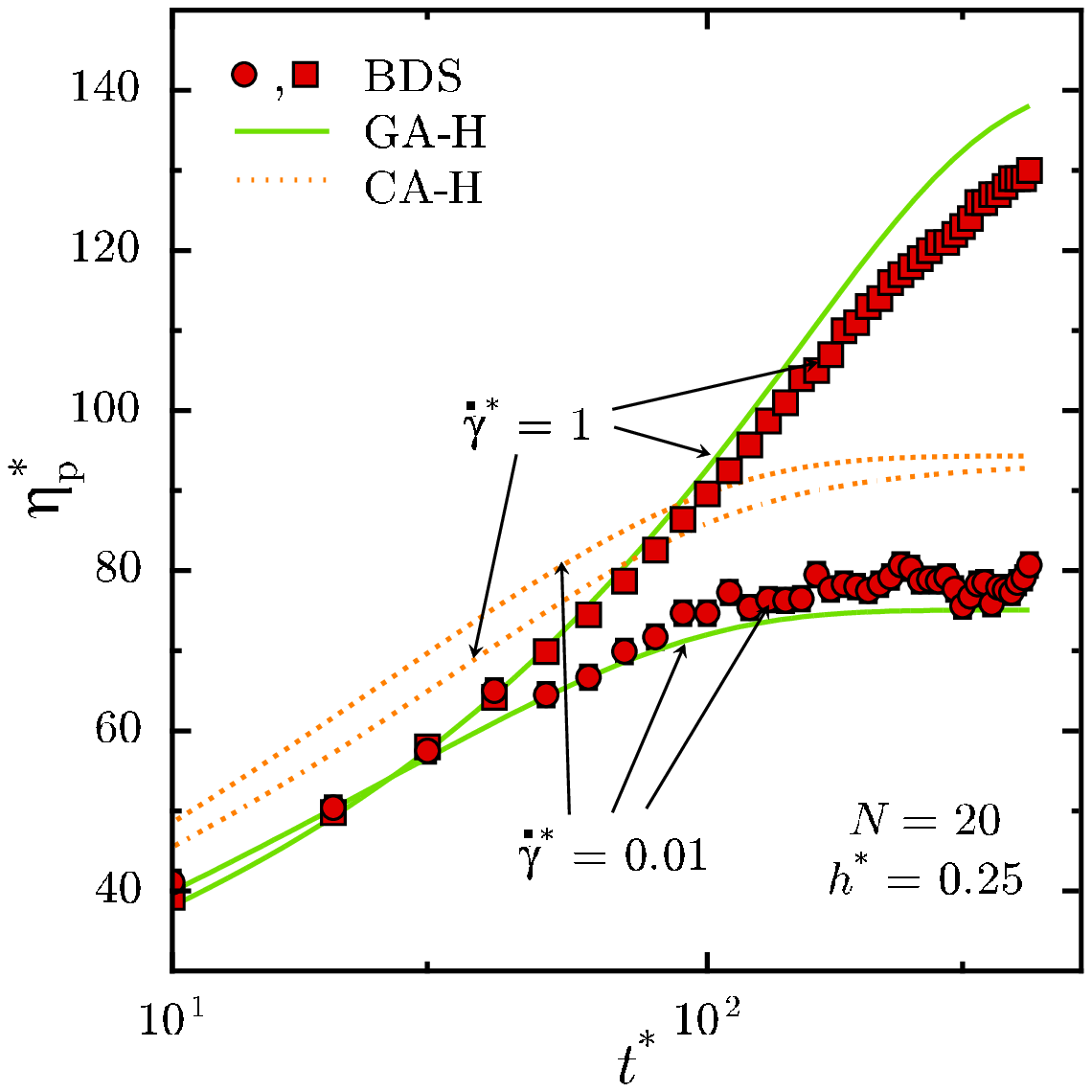}}}
\centerline {\resizebox{7cm}{!}{\includegraphics{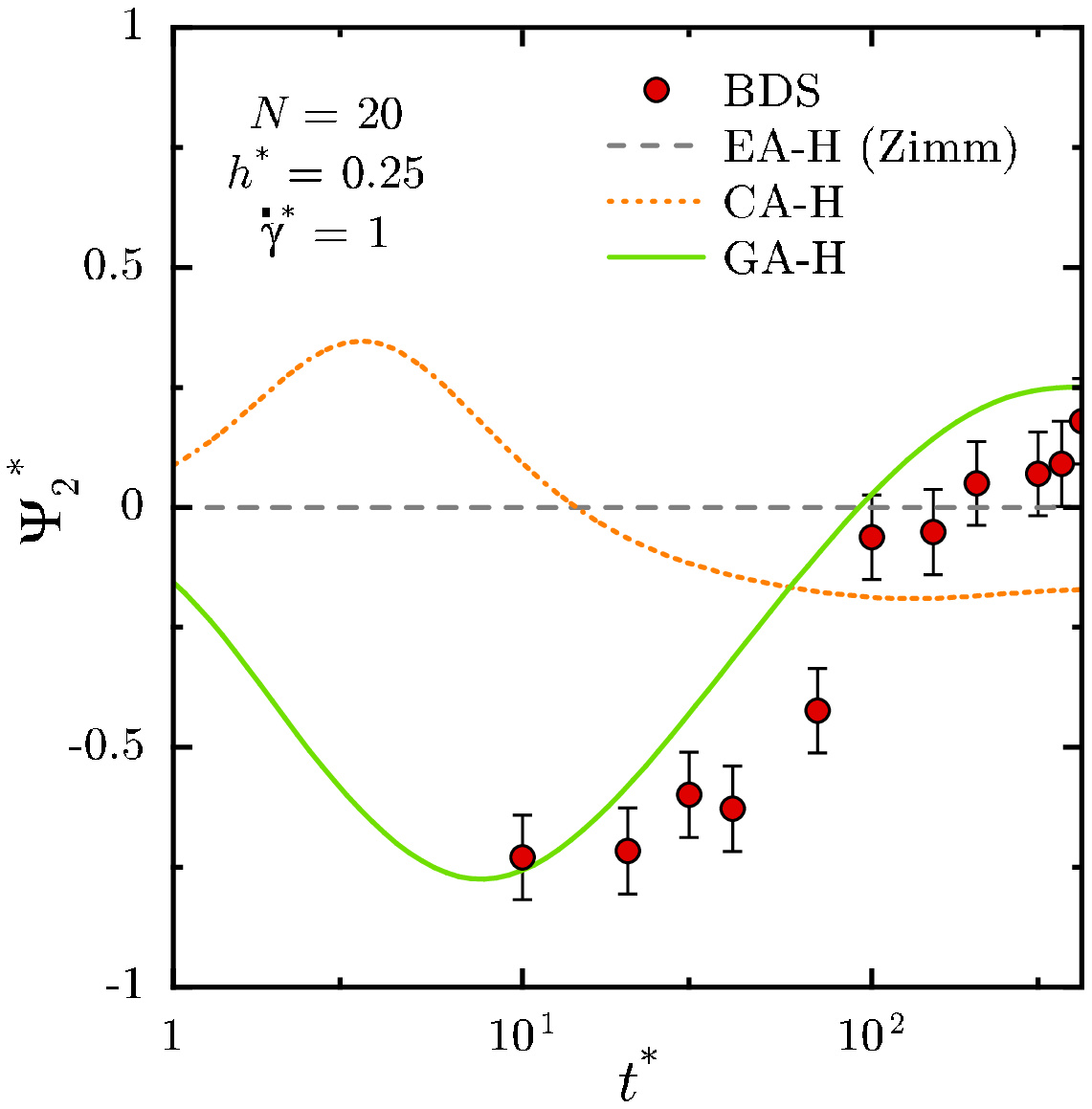}}}
\caption{\label{f:usetaHpsi2H} Growth of (a) viscosity, and (b)
second normal-stress difference coefficient, for Rouse chains with
HI during start-up of steady shear flow.}
\end{figure}

A striking feature is the large overshoots predicted by all the
models in the growth of $\etaps$ in Fig.~\ref{f:fdusetapsi2}~(a).
A similar prediction is obtained for $\PsiIs$. This feature is not
observed at any shear rate for Rouse chains with HI (see
Fig.~\ref{f:usetaHpsi2H}~(a)), and neither is it observed at low
shear rates for FEBS chains. The overshoots in the material
functions appear to be accompanied by a corresponding overshoot in
the mean-squared end-to-end distance of the coils. Overshoots in
the transient $\etaps$, $\PsiIs$ and $\Resqs$ during start-up of
shear flows are known to be a signature of the influence of FE
\citep{brule:jnnfm-93, herrchen:jnnfm-97, doyle:jnnfm-98a,
hur:jor-01}. \citeauthor{hur:jor-01} have obtained experimental
evidence of such overshoots using dilute (and semi-dilute)
solutions of DNA, and have shown using simulations of bead-rod and
bead-spring chains that the overshoot is caused by the rapid
localized stretching of the molecules by the flow, before the
orientation of the chains towards the flow direction reduces the
total solvent drag sufficiently to reduce the stretching. Herrchen
an d\"{O}ttinger\citep{herrchen:jnnfm-97} observed that the
overshoot predicted with FENE-P springs is larger than that
obtained with FENE springs in the case of dumbbells.

\begin{figure}[t]
\centerline {\resizebox{7cm}{!}{\includegraphics{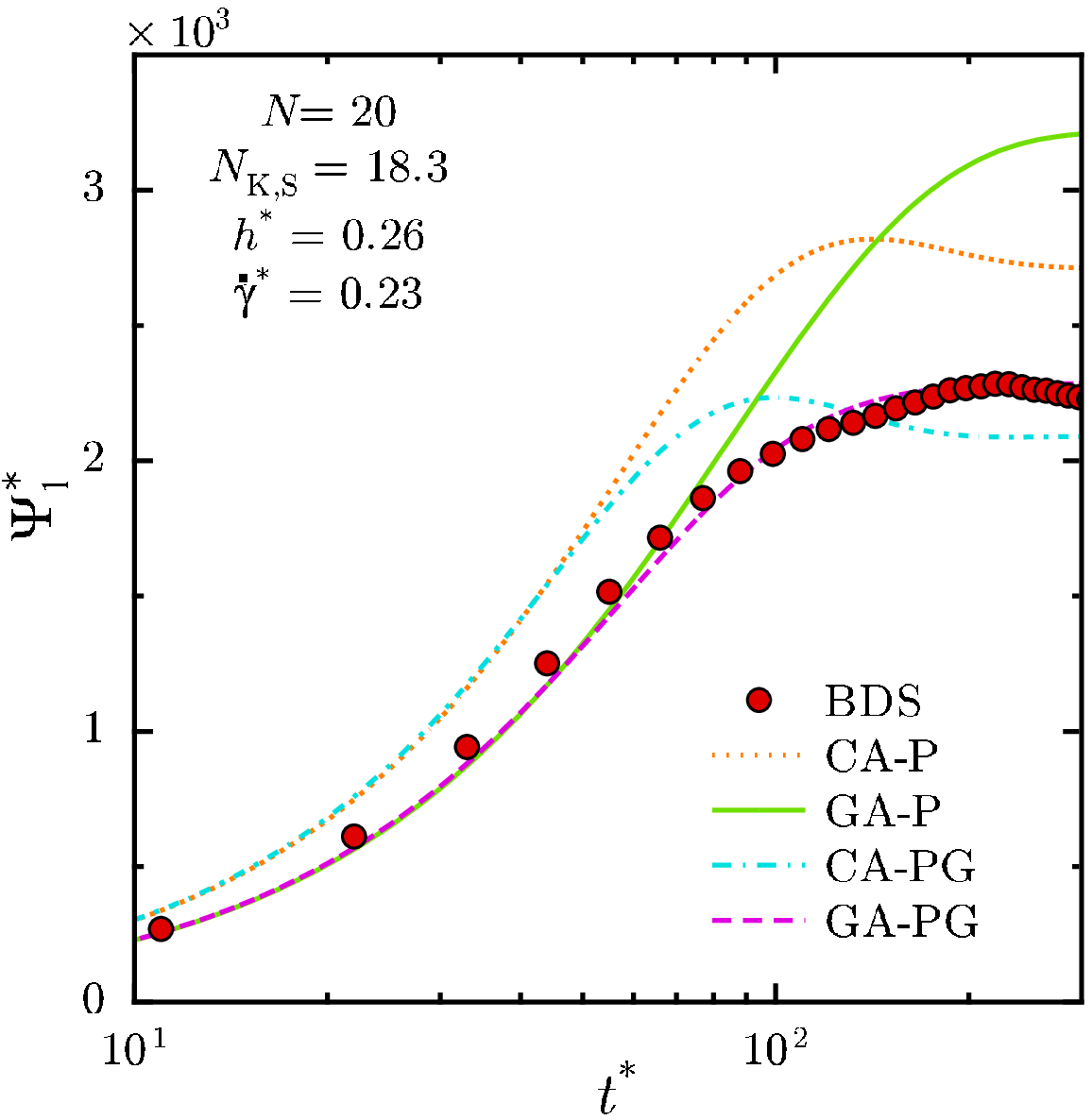}}}
\centerline {\resizebox{7cm}{!}{\includegraphics{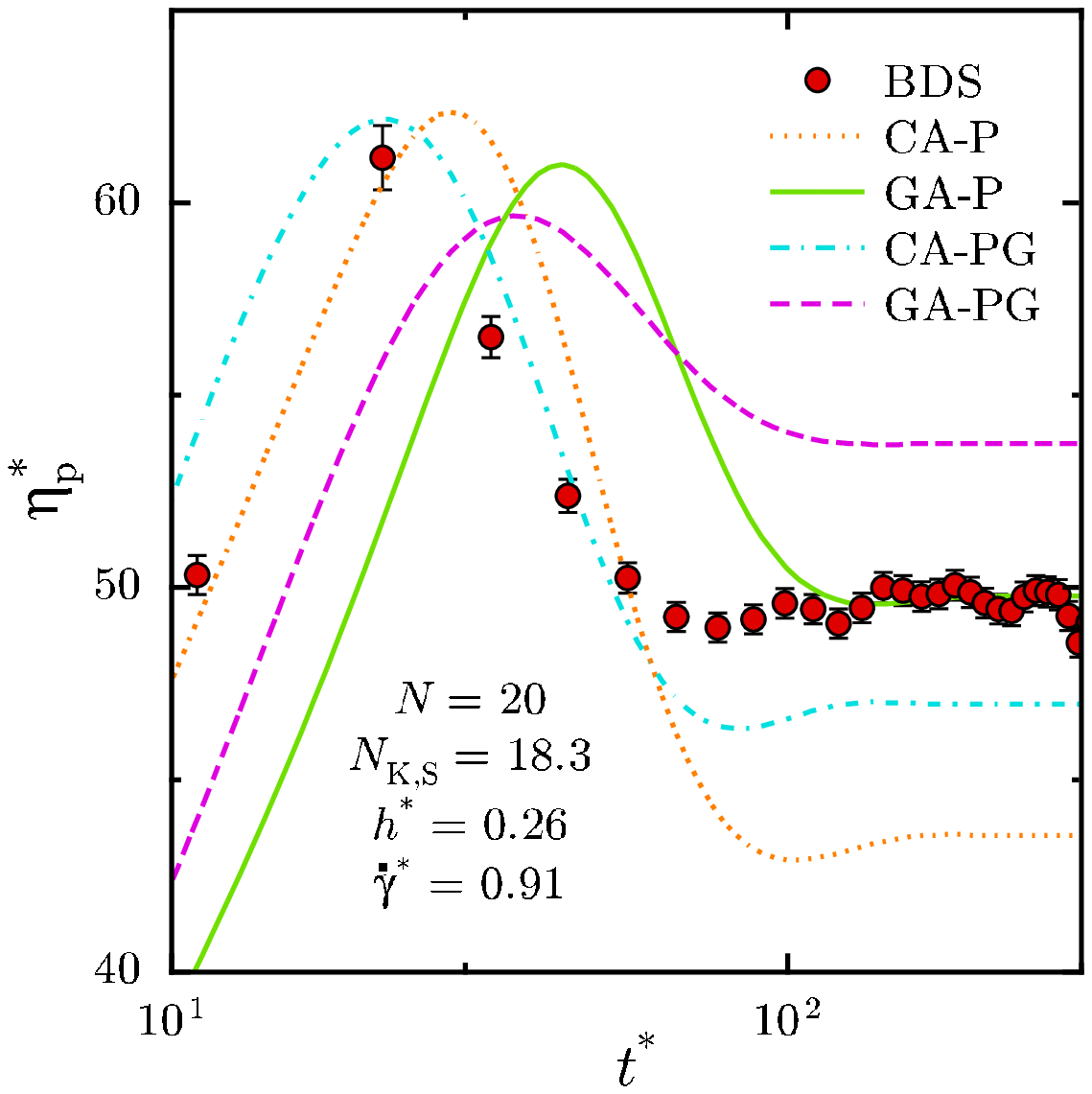}}}
\caption{\label{f:uslopsi1eta} Growth of (a) first normal-stress
difference coefficient, and (b) viscosity, for FEBS chains with HI
during start-up of steady shear flow.}
\end{figure}

Although the qualitative behaviour of the different approximations
for HI in Rouse chains has been well documented in the case of
steady shear flows, their predictions of rheological and
conformational properties in transient shear flows have not
received much attention. For Rouse chains with HI, use of
equilibrium averaging in the Zimm models results in a fixed time
dependence of properties that is independent of the shear rate
\citep{bird:dpl2}. This behaviour parallels the shear rate
independence of the material functions predicted at steady-state
by the Zimm model. In contrast, the predictions of the
consistent-averaging and Gaussian approximations, and the exact
results of the simulations for the growth of the shear material
functions depend on the imposed shear rate, as demonstrated in
Figs.~\ref{f:usetaHpsi2H}~(a) and~(b) for the transient variation
in $\etaps$ and $\PsiZs$, respectively, after the imposition of a
steady shear rate.

In Fig.~\ref{f:usetaHpsi2H}~(a), the curves predicted with the
Gaussian approximation initially lie below those predicted with
consistently averaged HI, for both the shear rates considered. At
larger values of $\ts$, as steady-state is approached, the
prediction of the Gaussian approximation at the higher shear rate
is seen to overtake the CA-H curve, whereas at the lower shear
rate, the GA-H prediction is always below that predicted by the
CA-H model.  The importance of including a description of the
fluctuations in approximate models accounting for HI is clearly
highlighted by the superiority of the predictions of the Gaussian
approximation. Interestingly, Fig.~\ref{f:usetaHpsi2H}~(b) shows
that fluctuations in HI lead to a negative second normal stress
difference coefficient soon after the inception of flow.
Zylka\citep{zylka:jcp-91} had earlier demonstrated  the accurate
prediction of the sign of $\PsiZs$ with the Gaussian approximation
at \textit{steady state}. The results here show further that the
approximation is excellent even in transient shear flows.

Figure~\ref{f:uslopsi1eta}~(a) shows that the GA-PG approximation
is particularly useful in shear flows of moderate strength where
fluctuations in the spring forces and HI are both important. The
prediction of the GA-PG model is in remarkable agreement with
simulations' results for the unsteady growth in the first
normal-stress difference coefficient following the imposition of a
moderate shear rate of $\gdots = 0.23$. Similar quantitative
agreement is obtained for the unsteady variation in $\etaps$ and
$\Resqs$ at this shear rate. At lower shear rates, FE is not
important, and the behaviour is the same as that obtained with
Rouse chains, for which it was shown above that the Gaussian
approximation does well in describing unsteady behaviour.

\begin{figure}[t]
\centerline{\resizebox{7cm}{!}{\includegraphics{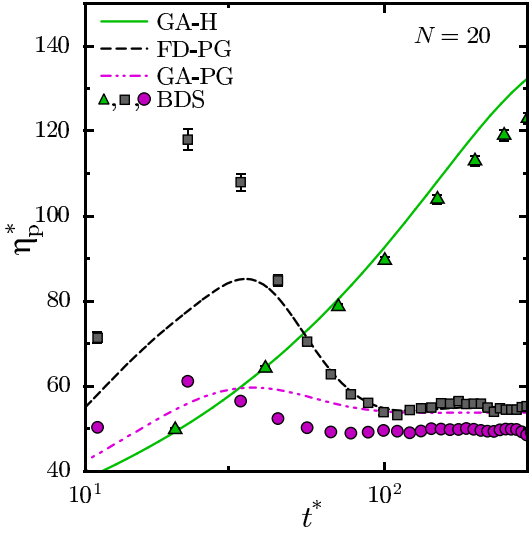}}}
\caption{\label{f:ushieta2} Comparison of the predictions of the
GA-H, FD-PG and GA-PG models for the growth of the shear viscosity
with BDS results obtained with Rouse chains with HI (green
triangles), free-draining FENE chains (grey squares), and FENE
chains with HI (magenta circles). The dimensionless shear rate is
unity for Rouse chains and 0.91 for FENE chains.}
\end{figure}

At high shear rates, where FE becomes the dominant effect,
however, predictions of no single model are uniformly in agreement
with the BDS results for all the properties examined, as
illustrated in Fig.~\ref{f:uslopsi1eta}~(b). To understand the
source of the deviations in the predictions of the GA-PG model, we
compare in  Fig.~\ref{f:ushieta2} predictions of the GA-H, FD-PG
and GA-PG models at $\gdots \sim 1$. The transient evolution of
the viscosity in the GA-PG model can be broken into two phases:
the early overshoot phase, and the later steady-state approach
phase. At early times, we see that the GA-H model is in very good
agreement with the BDS data (green triangles) for Rouse chains
with HI. In the same period, the FD-PG predictions deviate quite
strongly from the corresponding BDS data for free-draining FEBS
chains (grey squares). The agreement of the composite GA-PG model
with the BDS data for FEBS chains with HI (magenta circles) is
thus seen to be intermediate in the overshoot phase. To understand
the behaviour at steady state, we note that fluctuations in HI at
high shear rates aid in the penetration of the solvent's velocity
field into the deformed polymer coil, and the viscosity predicted
is higher and closer to predictions with the corresponding
free-draining model. This important aspect of HI's influence on
the behaviour of dilute polymer solutions will be more clearly
demonstrated shortly in predictions of steady state properties.
From the overprediction of the viscosity by the GA-H model at
large times in Fig.~\ref{f:ushieta2}, we see that  at high shear
rates, the Gaussian approximation for HI overestimates the
influence of fluctuations in HI. In Fig.~\ref{f:uslopsi1eta}~(b),
we see that the CA-PG model's prediction is lower than the BDS
results. The Gaussian approximation for HI thus pushes the
viscosity prediction of the CA-PG past the BDS results.

\subsubsection{\label{s:rsfb} Steady shear flow}

\begin{figure}[t]
\centerline {\resizebox{7cm}{!}{\includegraphics{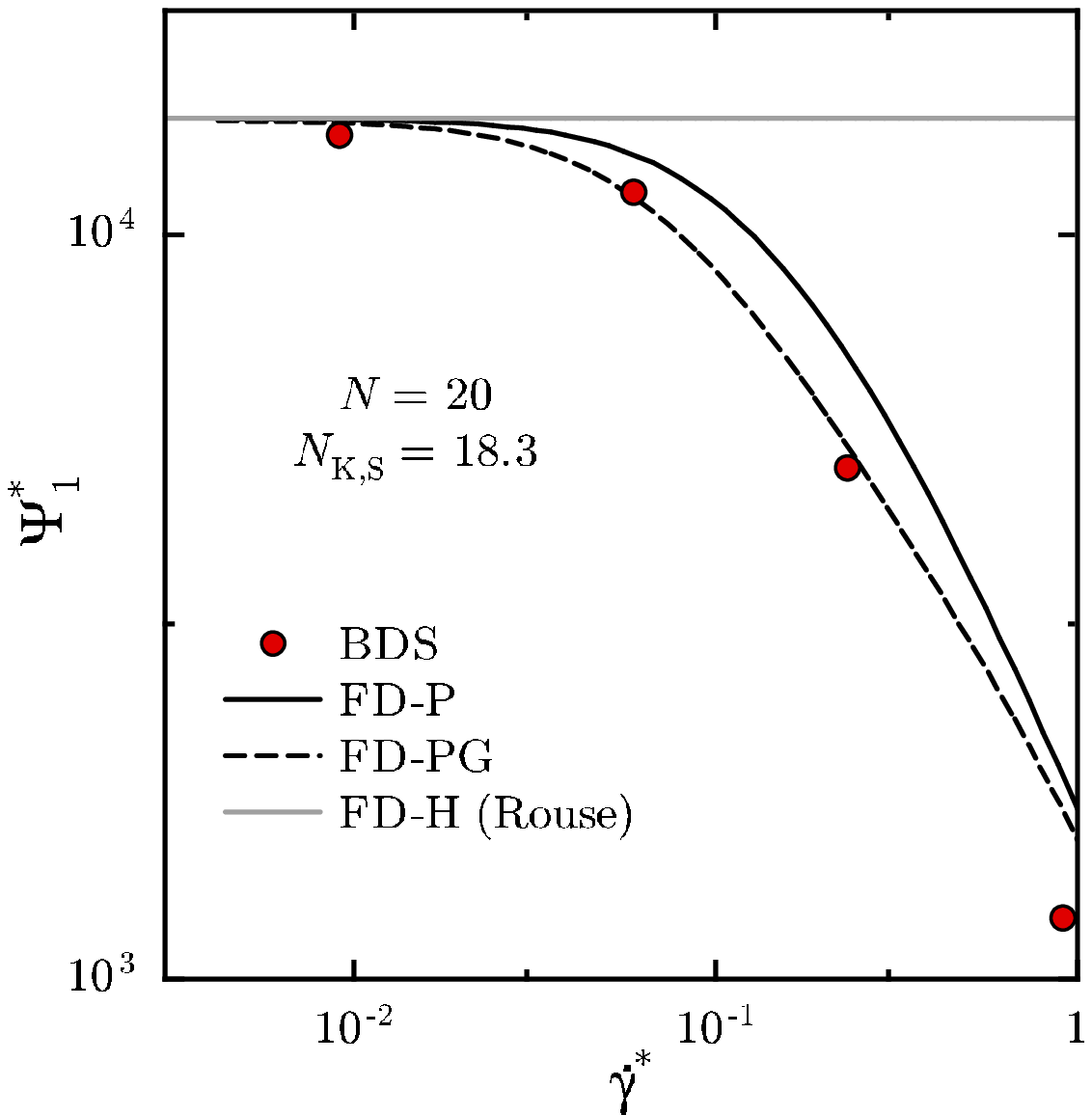}}}
\centerline {\resizebox{7cm}{!}{\includegraphics{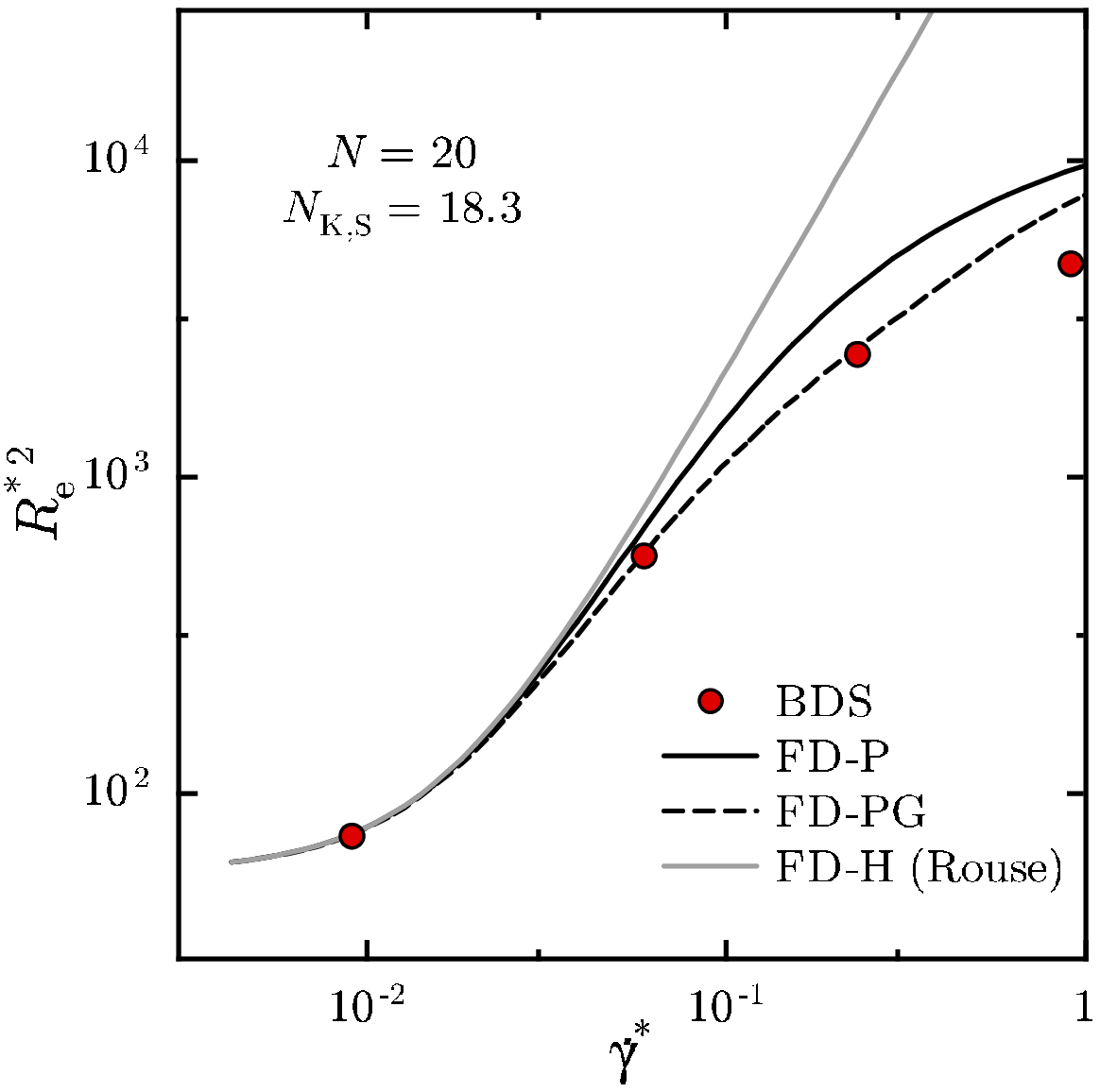}}}
\caption{\label{f:fdsspsi1} Variation of (a) the steady-state
first normal-stress difference coefficient, and (b) the
steady-state mean-squared end-to-end distance, with shear rate,
for free-draining FEBS chains.}
\end{figure}

The steady-state shear viscosity predicted by the FENE-P
approximation has been observed earlier to be in quantitative
agreement with the values obtained with BD simulations of small
free-draining FENE chains \citep{herrchen:jnnfm-97,
brule:jnnfm-93}. For chains with $N = 20$, it is observed that the
predictions of $\etaps$ obtained with the FENE-P and FENE-PG
approximations are nearly identical. However, in
Figs.~\ref{f:fdsspsi1}~(a) and~(b), it is seen that the FENE-PG
approximation for the FENE force law does much better than the
FENE-P approximation in its prediction of the steady-state
$\PsiIs$ and $\Resqs$, respectively, at moderate shear rates. At
$\gdots \approx 1$, the predictions of $\Resqs$ and $\PsiIs$ with
FENE-PG springs also begin to deviate from BDS results
significantly, but are closer to the exact results than the
predictions obtained with FENE-P springs. From the fact that the
stresses and $\Resqs$ values predicted with simulations and the
FD-PG model are lesser than the predictions of the FD-P model, we
conclude that fluctuations in the spring forces lead to an
increase in the effective stiffness of the chains.

\begin{figure}[t]
\centerline {\resizebox{7cm}{!}{\includegraphics{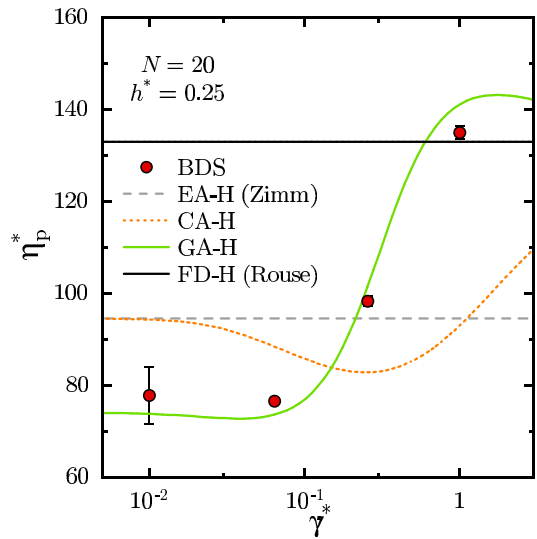}}}
\caption{\label{f:ssetaH} The influence of HI on the variation of
the steady-state polymer viscosity with dimensionless shear rate,
for Rouse chains. }
\end{figure}

As mentioned before, the predictions of the Gaussian approximation
for Rouse chains in steady shear flows have been compared earlier
with those of the Consistent- and Equilibrium-Averaging (Zimm)
approximations, and with the results of exact BD
simulations~\citep{ottinger:jcp-89, zylka:jcp-91}.
Figure~\ref{f:ssetaH} illustrates  some of the key features of the
predictions of the different approximations. As is well known, the
net reduction in the average solvent velocity gradient within the
polymeric coil \citep{yamakawa:mtps, larson:cepms, ottinger:ra-96}
caused by HI results in smaller values of the shear viscosity at
low to moderate shear rates, when compared to the free-draining
Rouse model. The fact that the predictions of the Gaussian
approximation and BD simulations  are lower in Fig.~\ref{f:ssetaH}
than those obtained with the Zimm and consistent-averaging
approximations near equilibrium and at moderate shear rates,
indicates that under these conditions, fluctuations in
configurations at equilibrium result in a more effective screening
of the velocity field within the coil. As the shear rate
increases, the steady-state viscosity and the first normal stress
difference coefficient first decrease in models with HI, other
than the Zimm model. This shear-thinning, although not very
strong, is a distinctive feature of configuration-dependent HI.
The initial shear-thinning is observed to be more prominent with
consistently-averaging than with the Gaussian approximation for
HI. As the shear rate increases, the predictions of $\etaps$ and
$\PsiIs$ obtained with BD simulations, and the CA-H and GA-H
models, reach a minimum with respect to the shear rate $\gdots$.
With a further increase in the shear rate, shear-thickening is
observed in these models, wherein $\etaps$ and $\PsiIs$ begin to
increase towards the constant value predicted by the Rouse model
in which HI is absent. The increase in $\etaps$ and $\PsiIs$
towards the Rouse model's predictions is understood to be the
consequence of the decreasing influence of HI as the average
separation between the beads in the chains increases.
Interestingly, in Fig.~\ref{f:ssetaH}, the predictions of the GA-H
model and the BD simulations' results lie above the curve
predicted by the CA-H model beyond a threshold shear rate, and are
closer to the free-draining Rouse model's prediction. This
behaviour is contrary to that observed near equilibrium, and shows
that the role played by fluctuations in HI reverses at higher
shear rates, and in fact aid the penetration of the solvent
velocity field into the polymer coil as the coil deforms and
becomes highly anisotropic.

It is known that both the consistent-averaging and Gaussian
approximations lead to shear rate dependent predictions for
$\PsiZs$, which are opposite in sign for a large range of shear
rates. This was first shown by Zylka\citep{zylka:jcp-91}, thus
demonstrating that fluctuations in HI lead to a change in the sign
of $\PsiZs$ at low to moderate shear rates for Rouse chains,
inline with the results obtained here above for unsteady shear
flow.

\begin{figure}[t]
\centerline {\resizebox{7cm}{!}{\includegraphics{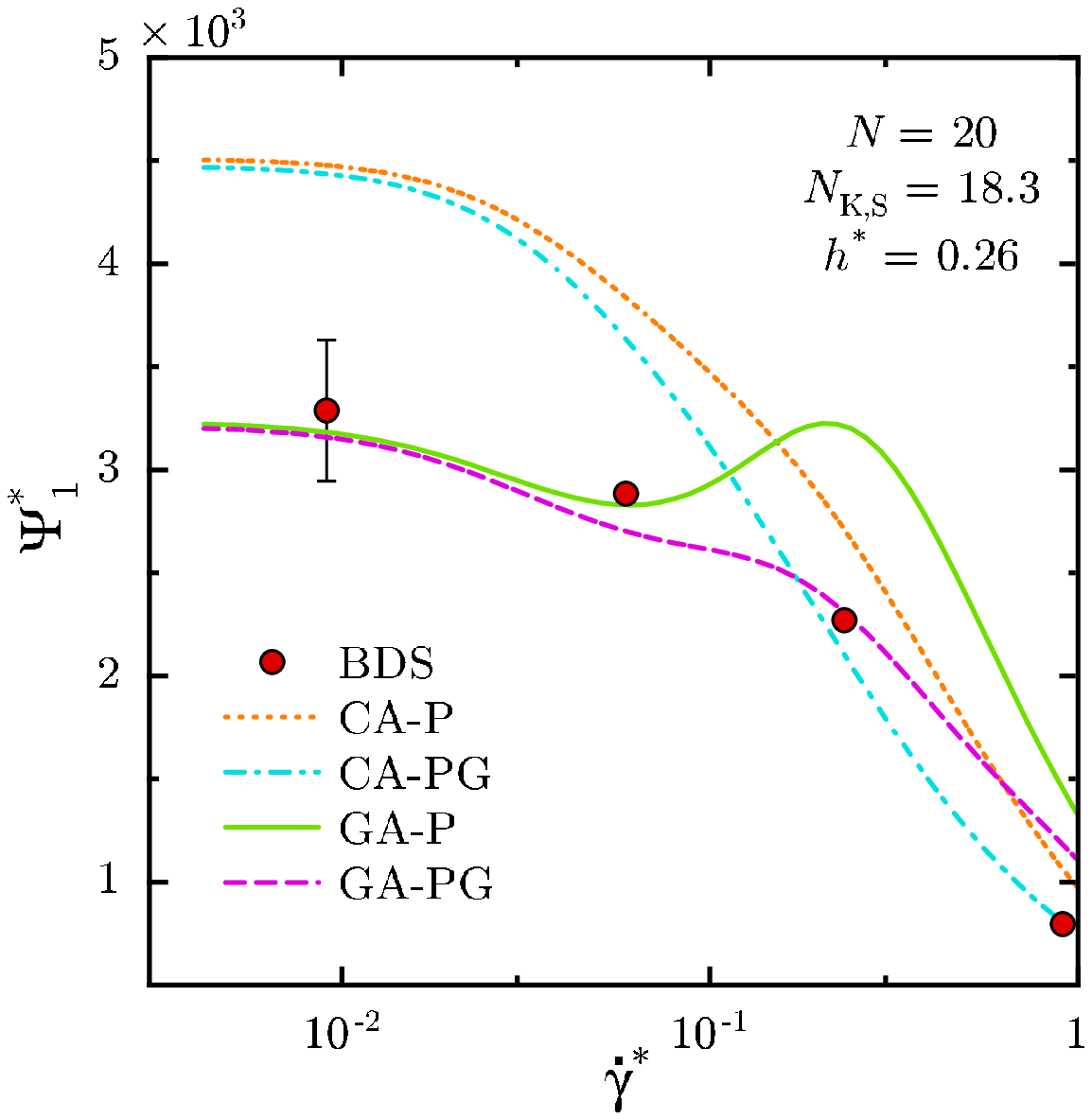}}}
\caption{\label{f:sspsi1} Variation of the steady-state first
normal-stress difference coefficient with shear rate, for FEBS
chains with HI.} \centerline
{\resizebox{7cm}{!}{\includegraphics{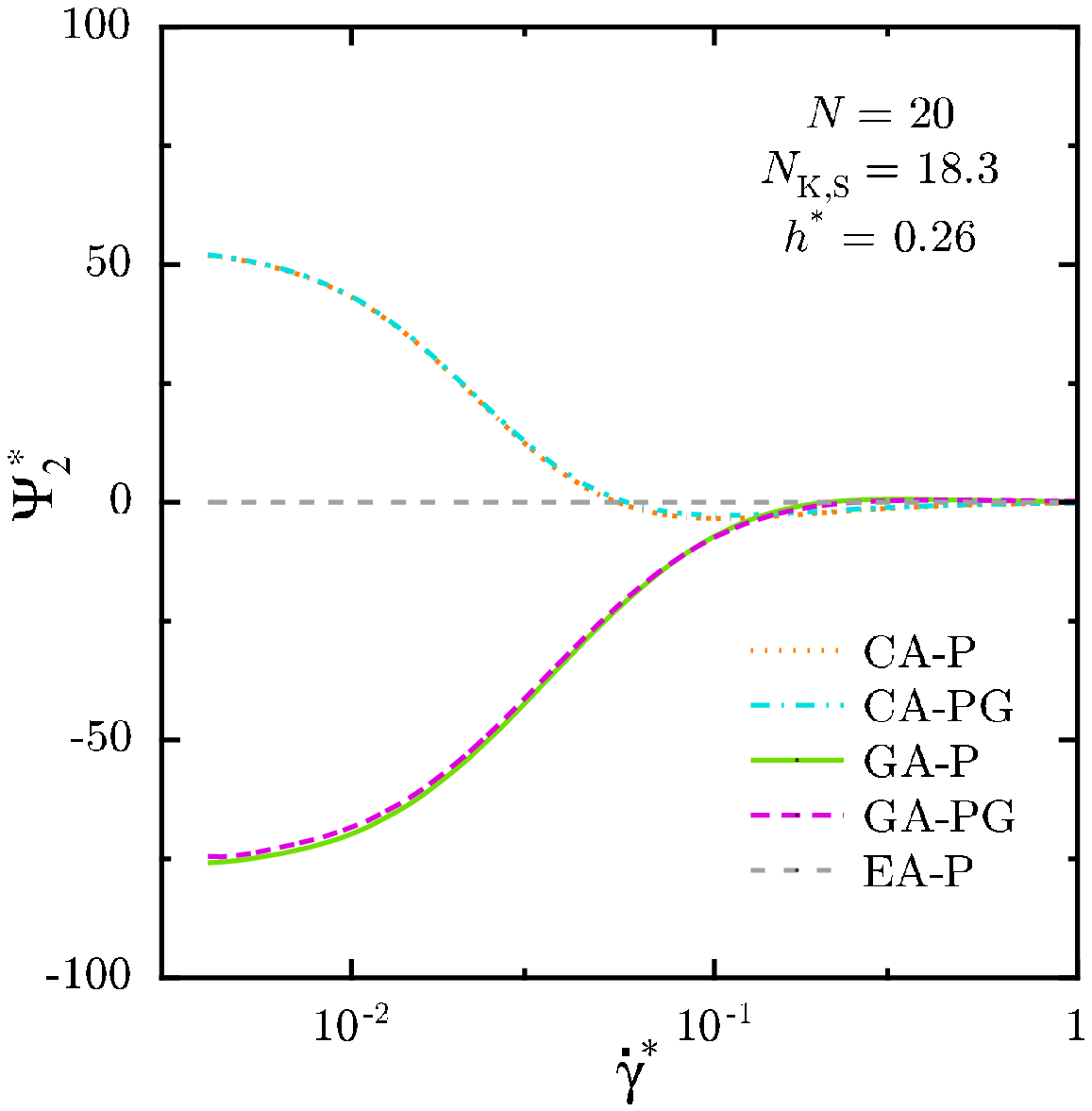}}}
\caption{\label{f:sspsi2} Variation of the steady-state second
normal-stress difference coefficient with  shear rate, for FEBS
chains with HI.}
\end{figure}

Figure \ref{f:sspsi1} shows the results for the steady-state
$\PsiIs$ obtained with the different approximate models combining
HI and FE. Predictions for the transient growth of $\PsiIs$ at
$\gdots = 0.23$ were presented earlier in Fig.~\ref{f:uslopsi1eta}
(a). The qualitative features of the predictions for the
steady-state viscosity $\etaps$ are the same, although the
shear-thinning-thickening behaviour is more pronounced in the case
of $\PsiIs$.  As before, the predictions of these models are
compared with the results of BD simulations of FENE chains with
fluctuating HI.

It is clear from Fig.~\ref{f:sspsi1} that among all the
approximate models, only the predictions of the GA-PG model are
close to the simulations' results across the whole range of shear
rates examined. As one would expect, at low shear rates where the
influence of FE is negligible, the observed behaviour is close to
that observed for Rouse chains with HI, for which the Gaussian
approximation for HI has been shown earlier to perform better than
the Consistent-Averaging approximation. At moderate shear rates
when FE begins to exert its influence, it is seen that the
Gaussian approximation is not as accurate when used with FENE-P
springs as it is with FENE-PG springs. This suggests that
fluctuations in the spring forces also become important at
moderately large shear rates.  Interestingly, spring force
fluctuations are however observed in Fig.~\ref{f:sspsi2} to have a
negligible effect on the steady-state $\PsiZs$, and the
predictions with the FENE-PG force expression are nearly the
identical to those obtained with the FENE-P approximation.

As the shear rate increases further, the differences between the
consistent-averaging and Gaussian approximations reduce for the
same choice of the spring force law. However, it is observed
\citep{prabhakar:phd-05} that at high shear rates, the predictions
with models using the Gaussian Approximation for HI are closer to
their free-draining counterparts, than those using
consistently-averaged HI. This behaviour once again points to the
fact that fluctuations in HI aid in the penetration of the
solvent's velocity gradient into the deformed polymer coil at
higher shear rates.

\begin{figure}[t]
\centerline{\resizebox{7cm}{!}{\includegraphics{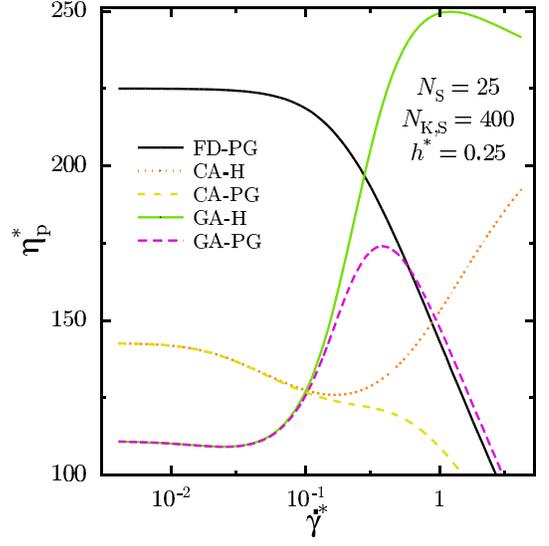}}}
\caption{\label{f:thikthin1} Prediction of
shear-thinning-thickening-thinning of the steady-state shear
viscosity by models with HI, for $\NK = 10^4$. }
\end{figure}

Although at high shear rates Rouse chains with HI predict
shear-thickening, FEBS models with HI predict shear-thinning.
Kisbaugh and McHugh\citep{kisbaugh:jnnfm-90} showed using the
diagonalized version of the CA-P model that complicated
shear-tinning-thickening-thinning behaviour can occur as a result
of the interplay between the emergence of FE and the fading of HI.
Figure~\ref{f:thikthin1} shows the predictions of $\etaps$
obtained with the CA-PG and GA-PG models for chains with 25
springs, and $\NKS = 400$. The behaviour in Fig.~\ref{f:thikthin1}
suggests that the curve for the steady-state $\etaps$ predicted by
a model combining the Gaussian Approximation for HI and the
FENE-PG approximation for the nonlinear spring force is
approximately bounded from above by the envelope formed by the
Gaussian Approximation for Rouse chains at low to moderate shear
rates, and  by  the terminal shear-thinning curve predicted by the
free-draining model with the FENE-PG springs at high shear rates.
The shear rate at which the $\etaps$-versus-$\gdots$ curves
predicted by the two simpler models (\emph{viz.} the GA-H and
FD-PG models) intersect can be regarded as a critical shear rate
$\gdots_\textrm{c}$. At shear rates smaller than
$\gdots_\textrm{c}$, the overall behaviour of the polymer chains
is dominated by HI, whereas FE is the dominant phenomenon when
$\gdots > \gdots_\textrm{c}$.

\subsection{\label{s:ref} Extensional flow}

\subsubsection{\label{s:refa} Start-up and cessation of steady
extensional flow}

Models using Hookean springs predict unbounded extension is strong
extensional flows. As a result of this unrealistic behaviour, the
predictions of approximations for HI in Rouse chains have not been
examined in strong and unsteady extensional flows. However, at any
finite time following the imposition of the flow, the values of
the stresses and other properties obtained with Rouse chains are
still finite. Therefore, by comparing the predictions of the
approximations and the results of BD simulations for Rouse chains
one can study in isolation the influence of HI in strong
extensional flows.

\begin{figure}[t]
\centerline {\resizebox{7cm}{!}{\includegraphics{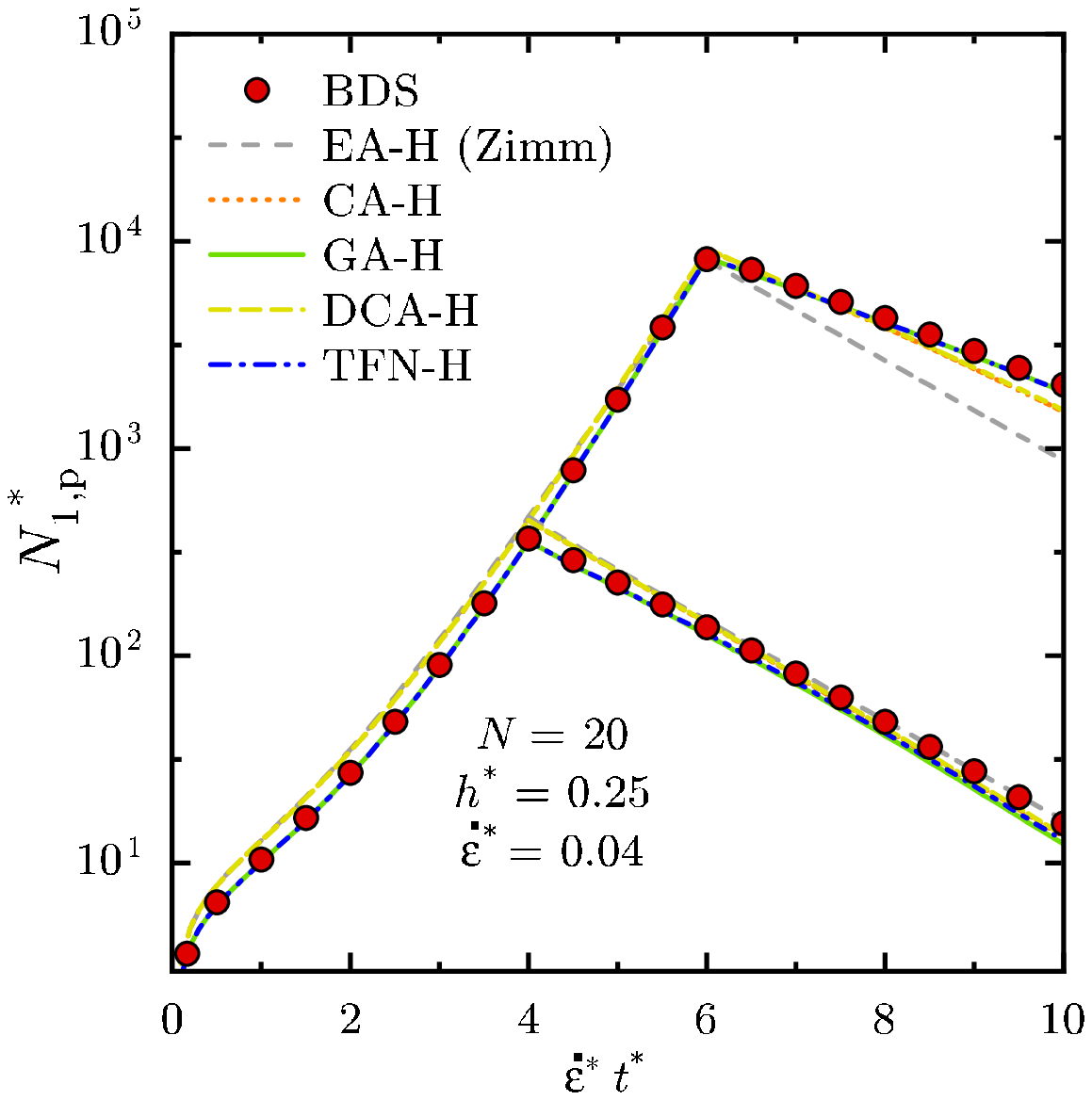}}}
\caption{\label{f:uun1H} Growth and decay of the polymer
contribution to the first normal-stress difference for Rouse
chains with HI, during start-up and following cessation of steady
extensional flow, respectively. }

\centerline {\resizebox{7cm}{!}{\includegraphics{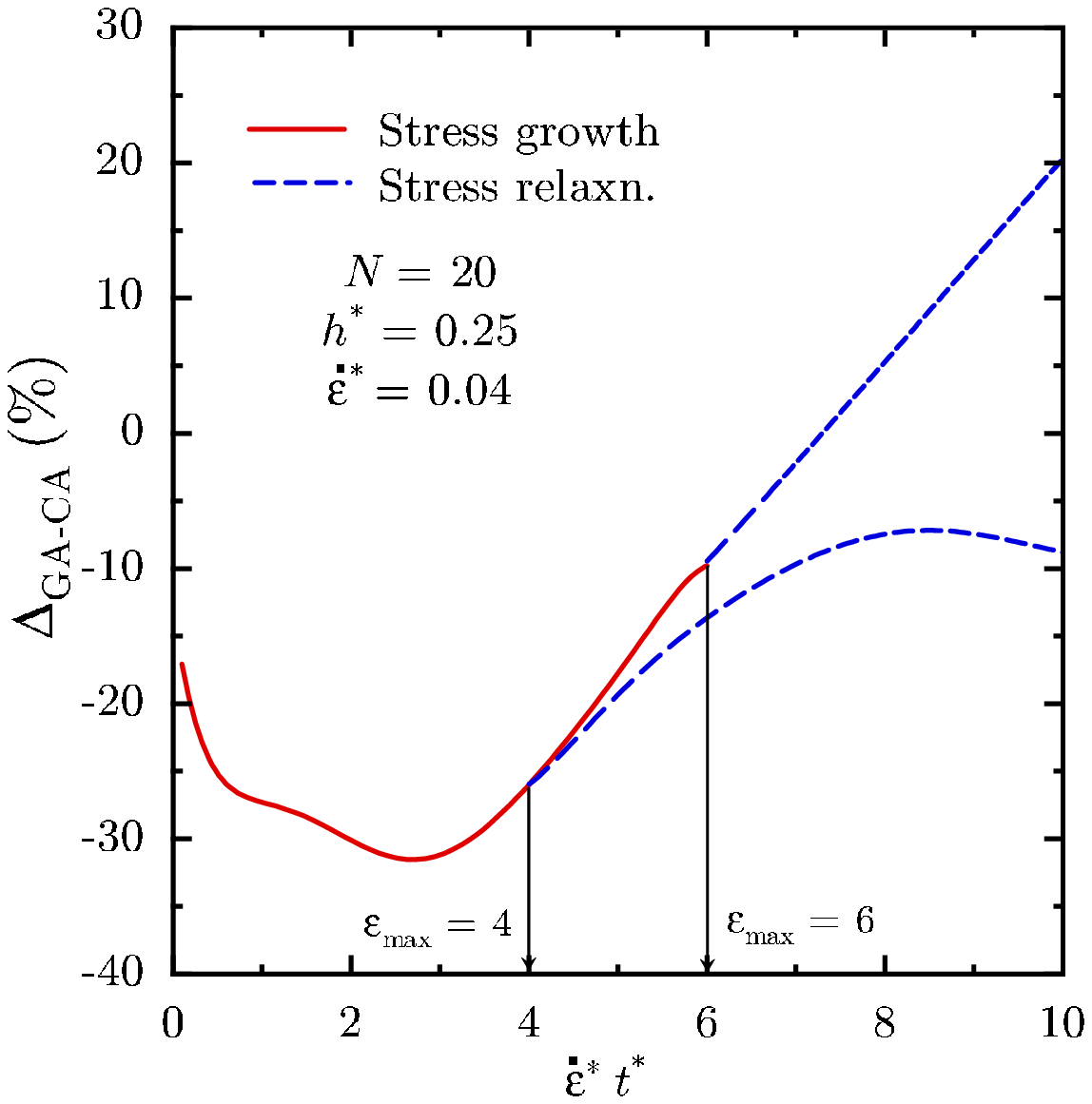}}}
\caption{\label{f:uucagadH} Effect of fluctuations in HI on the
growth and decay of the polymer contribution to the first
normal-stress difference for Rouse chains with HI, during start-up
and following cessation of steady extensional flow, respectively.}
\end{figure}

Figure~\ref{f:uun1H} shows the growth of $\NIps$ upon the
imposition of a steady extensional flow of $\edots = 0.04$ (which
is greater than the critical strain rate for the coil-stretch
transition $\edotsc = 0.012$ for $N = 20$ and $\hs = 0.25$). Also
shown in Fig.~\ref{f:uun1H} are the predictions for the relaxation
of $\NIps$ after the cessation of the extensional flow at
different values of the Hencky strain.

The results in Fig.~\ref{f:uun1H} show several fascinating
features. Firstly, the predictions of all the approximate models,
including the Zimm model with equilibrium-averaged HI appear to be
in remarkably close agreement with the exact results of BDS in the
stress growth phase. The Zimm model only begins to deviate
noticeably from the rest of the data during stress relaxation
after flow stoppage at high strains. Similar results are also
obtained  for $\Resqs$ \citep{prabhakar:phd-05}. Although the
differences between the different approximations appear to be
small on the scale of the plot in Fig.~\ref{f:uun1H}, in
Figure~\ref{f:uucagadH}, the differences between the predictions
of these approximations are examined more closely by plotting the
relative difference between the predictions of $\NIps$ of the two
approximations,
\begin{gather}
\Delta_\textsc{ga-ca} = \frac{{\NIps}_\textsc{ga} -
{\NIps}_\textsc{ca}}{{\NIps}_\textsc{ga}}\,.
\end{gather}
It is seen that fluctuations in HI can account for as much as 30\%
of the polymer stress in extensional flows.


\begin{figure}[t]
\centerline {\resizebox{7cm}{!}{\includegraphics{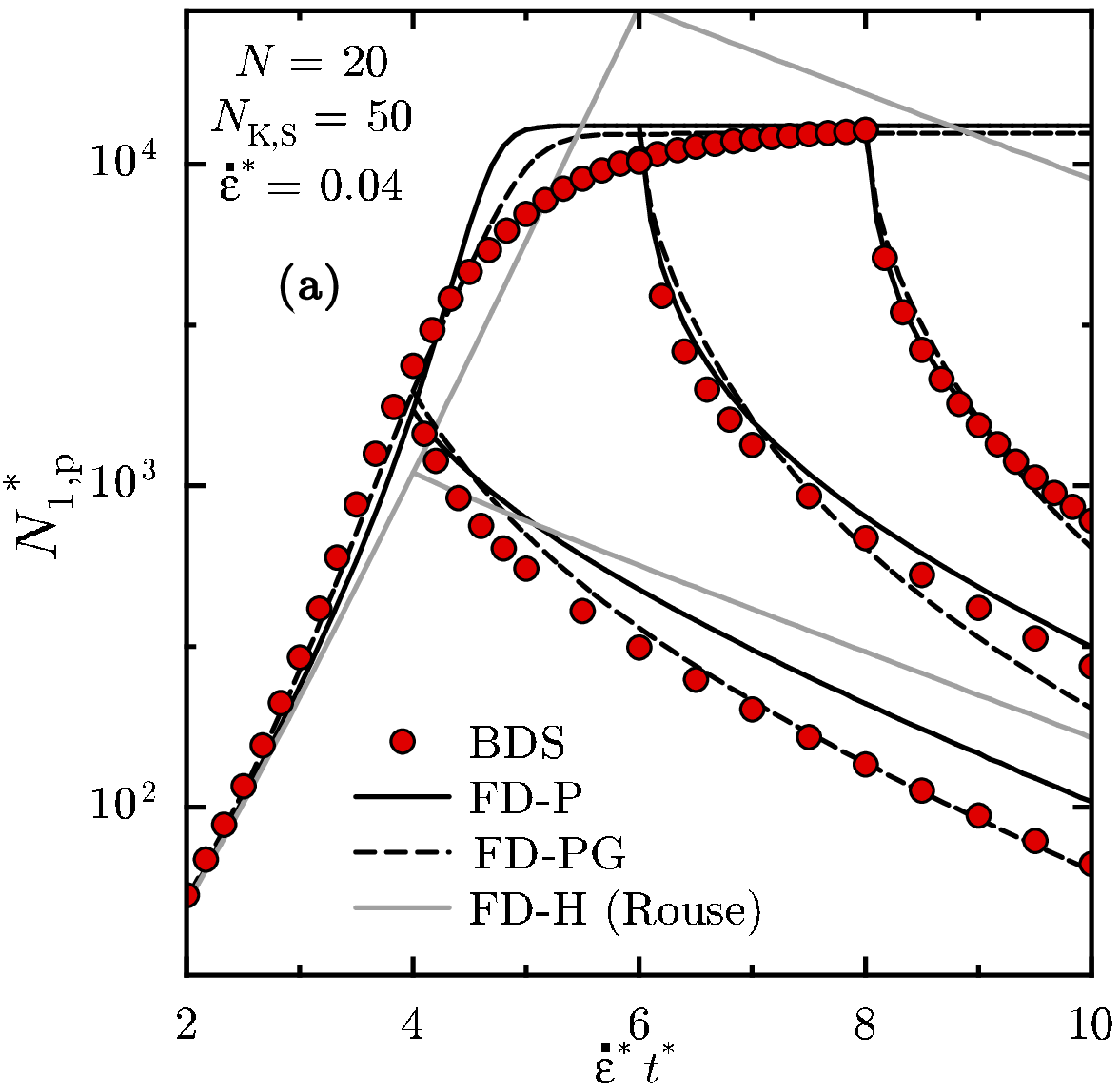}}}
\centerline{\resizebox{7cm}{!}{\includegraphics{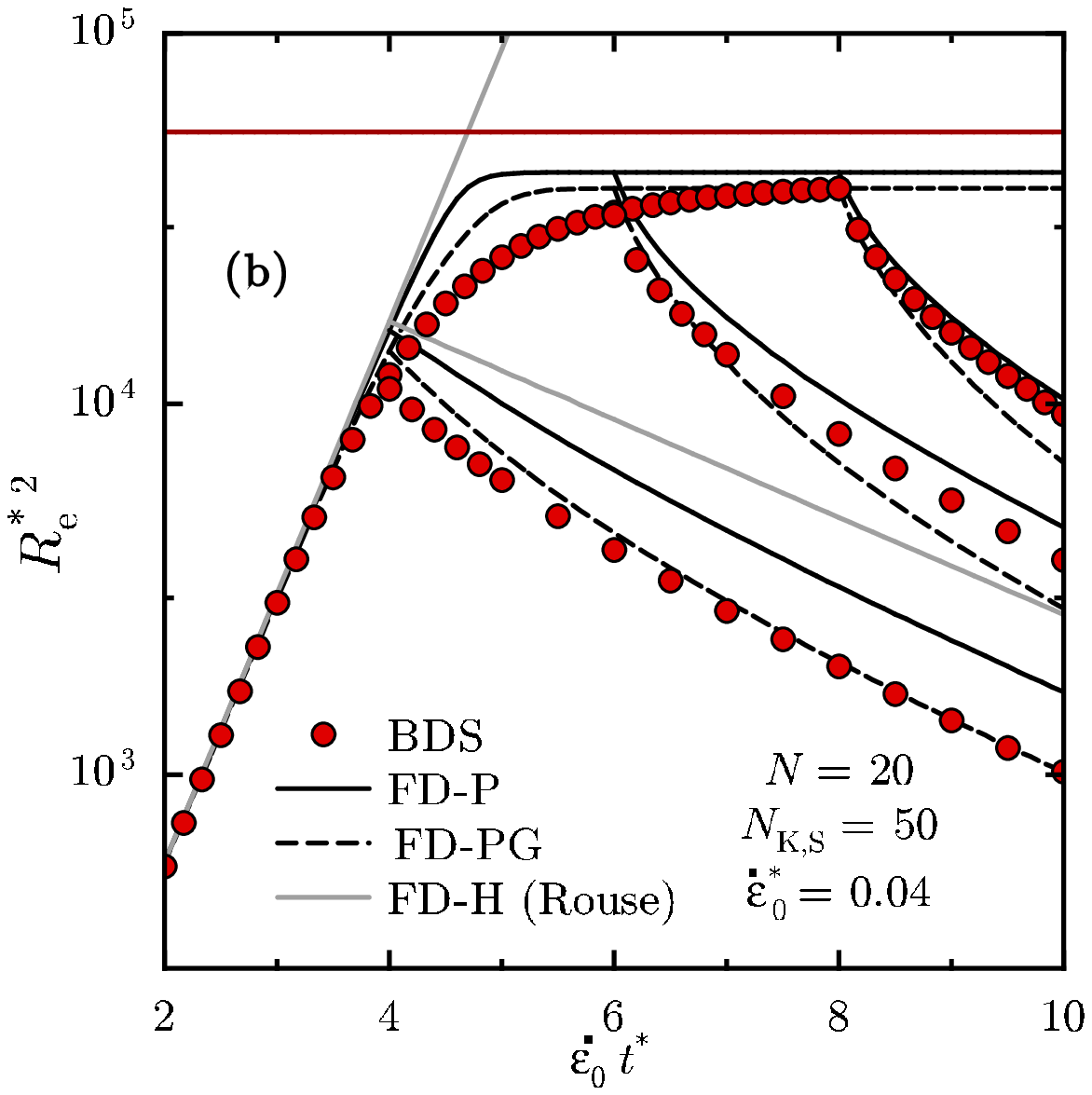}}}
\caption{\label{f:fduuN1lo} Growth of (a) the polymer contribution
to the first normal-stress difference, and (b) the mean-squared
end-to-end distance of free-draining FEBS chains, during start-up,
and their relaxation following cessation, of steady extensional
flow. The horizontal red line in (b) indicates  the maximum
possible value of $\Resqs = 3 \NKS \NS^2$.}
\end{figure}

Predictions of free-draining FEBS models employing the FENE-P and
FENE-PG approximations are compared with exact results of the
simulations for the growth and relaxation of $\NIps$ and $\Resqs$
in Figs. \ref{f:fduuN1lo} (a) and (b), respectively. Also shown
for comparison are the predictions of the simple Rouse model. The
strain-rate for which the data are shown in Fig.~\ref{f:fduuN1lo}
is larger than the critical strain rate $\edotsc = \sin^2 (\pi/2N)
= 6.2 \times 10^{-3}$ for free-draining chains with $N = 20$. At
this relatively high strain rate, the Rouse model predicts an
exponential increase in both $\NIps$ and $\Resqs$ during
extension. At small values of the Hencky strain during start-up of
the extensional flow at this strain rate, all the data for the
different models are essentially identical, since the springs are
not stretched sufficiently for the spring-force nonlinearities to
have any significant effects. The influence of FE is first
discerned in the growth of the stress when the curves for the FEBS
chains separate from the prediction of the Rouse model. At higher
strains, the deviations between the approximations and the exact
results for $\NIps$ become more marked and reach a maximum when
the approximate results level off sharply towards their eventual
steady-states. The approach of the BD simulations' data to
steady-state is more gradual. In the stress growth phase, the
predictions of the FENE-PG approximation are closer to the exact
results, than those obtained with the FENE-P approximation.

The relaxation behaviour of free-draining FEBS chains following
the cessation of extensional flow has been previously studied
\citep{doyle:jnnfm-98b, wiest:polymer-99}. While
\citeauthor{doyle:jnnfm-98b} compared the predictions of the
FENE-PM approximation  with BD simulations of FENE dumbbells and
also bead-rod chains, \citeauthor{wiest:polymer-99} examined some
interesting qualitative features in the stress-relaxation phase
using the FENE-P model.

The FENE-PG approximation is seen to lead to more accurate
description of both the stress and extension in the relaxation
phase when the strain at flow stoppage is not much larger than the
strain at which FE begins to exerts its influence, whereas the
FENE-P approximation leads to an overprediction of both
properties. When the Hencky strain at flow stoppage is larger, the
FENE-PG force law is seen to be in fact less trustworthy. This
loss in accuracy of the FENE-PG model when chains are highly
stretched is consistent with the behaviour observed earlier in
steady and unsteady shear flow.

\begin{figure}[t]
\centerline {\resizebox{7cm}{!}{\includegraphics{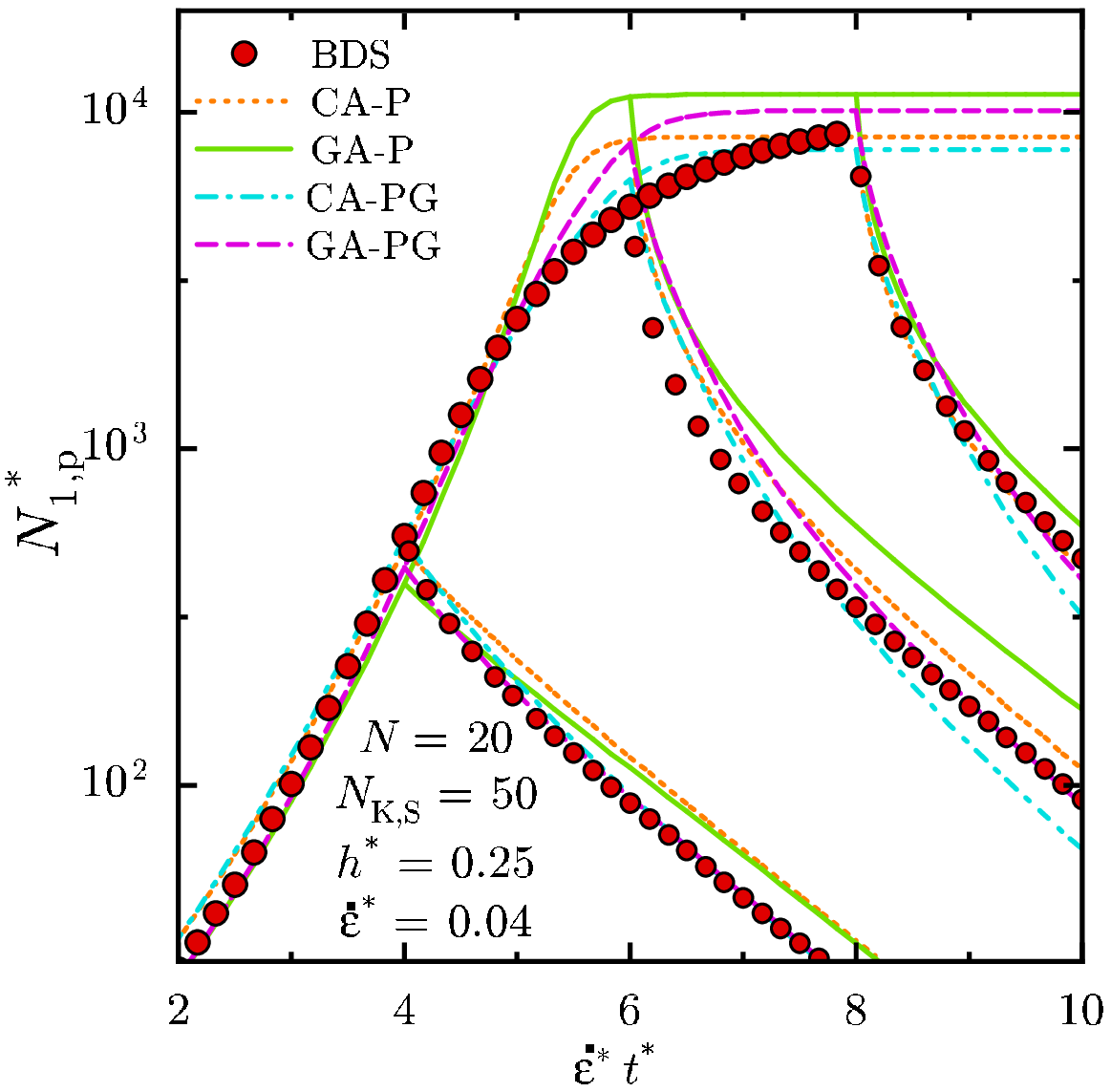}}}
\centerline {\resizebox{7cm}{!}{\includegraphics{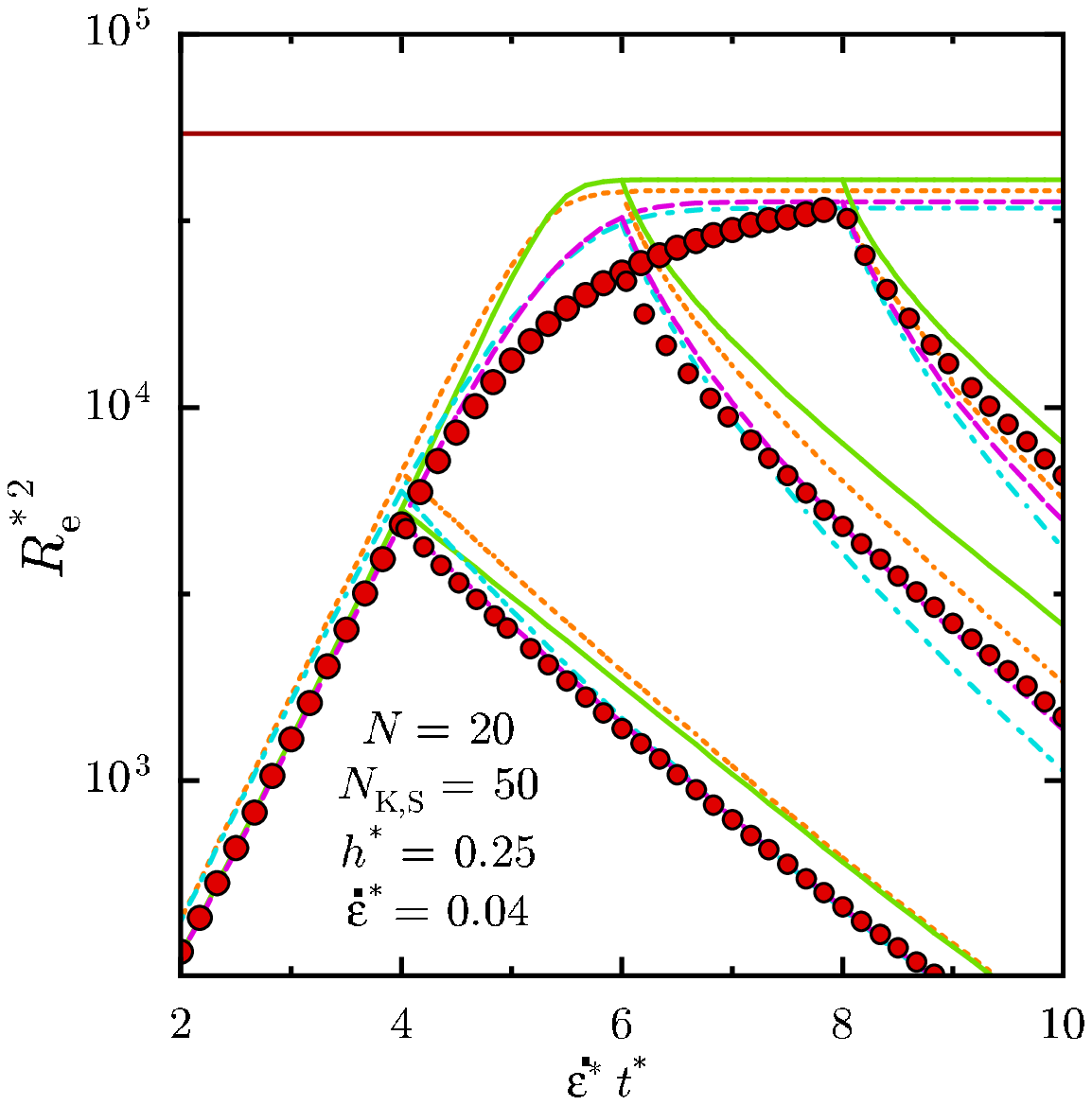}}}
\caption{\label{f:uun1} Growth and decay of the polymer
contribution to (a) the dimensionless first normal-stress
difference, and (b) the mean-squared end-to-end distance, for FEBS
chains with HI, during start-up and following cessation, of steady
extensional flow. The horizontal red line indicates the maximum
possible value of $\Resqs = 3\, \NKS \, \NS^2$.}
\end{figure}

The predictions of approximate models with both nonlinear
phenomena incorporated, for the unsteady growth and relaxation of
$\NIps$ and $\Resqs$ following start-up and cessation of an
extensional flow are shown in Figs.~\ref{f:uun1}~(a) and~(b). The
initial behaviour ($\He \leq 4$) in the stress-growth phase is
close to that observed for Rouse chains with HI since the
influence of FE is weak at these strains. The differences between
the various approximations grow larger at intermediate strains
during extension. At these strains and above, the curves predicted
by the combined models are qualitatively similar to the
predictions obtained with the approximations for free-draining
FEBS chains. In particular, the nature of deviations between the
BD simulations' results and the approximations' predictions appear
to be decided largely by the approximation used for the FENE
nonlinearity. At higher strains, as chain become highly stretched,
incorporating fluctuations in HI through the Gaussian
Approximation make the curves predicted by the combined models lie
closer to those obtained with the corresponding free-draining
model \citep{prabhakar:phd-05}. Predictions of the relaxation
behaviour are observed to be better with the FENE-PG force law
when the strain at flow stoppage is low. The largest deviations in
the approximations' predictions of the relaxation behaviour occur
when flow is stopped at a strain in the ``knee" region just prior
to steady-state. On the whole, the GA-PG model appears to be quite
accurate in its description of the normal stresses and chain
deformation in unsteady extensional flows.

The prediction of stress-conformation hysteresis is another test
of the accuracy of closure approximations. Stress-conformation
hysteresis in dilute polymer solutions has been studied in
experiments \citep{spiegelberg:icr-96, orr:jnnfm-99} and
theoretically \citep{doyle:jnnfm-98b, lielens:jnnfm-98,
wiest:polymer-99, li:ra-00, ghosh:jor-01}. In extensional flow
experiments, it is observed that when the normal stress is plotted
against the birefringence of the solution, the curves described by
the data points during the stress-growth and stress-relaxation
phases are very different. In particular, at any fixed value of
the birefringence, the stress during extension is larger than that
measured during relaxation. Conversely, at a fixed stress, the
birefringence during relaxation is larger.

\begin{figure}[!t]
\centerline {\resizebox{7cm}{!}{\includegraphics{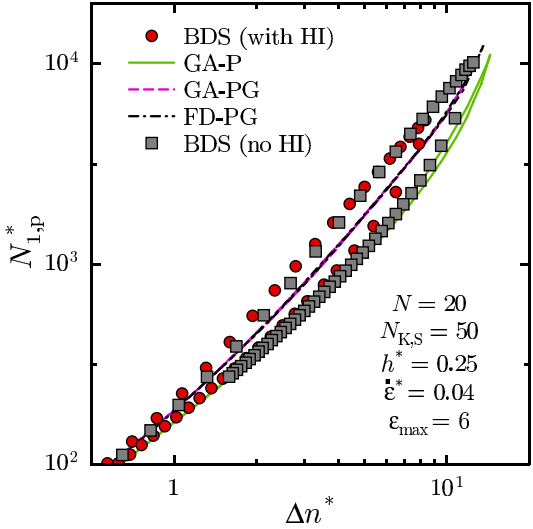}}}
\caption{\label{f:uun1dn} Stress-conformation hysteresis for FEBS
chains with and without HI.}
\end{figure}

Figure~\ref{f:uun1dn} compares the $\NIps$-versus-$\Delta n^\ast$
behaviour predicted by BD simulations and the various
approximations. It is immediately clear that closure
approximations fail to reproduce the size of the hysteresis loops
observed in the simulations. Models using the FENE-P springs
predict larger birefringence than the exact simulations at any
level of the stress. When fluctuations are accounted for in the
spring forces through the FENE-PG force law, the size of the
hysteresis loop becomes even smaller. Nevertheless, the curves
predicted with FENE-PG springs now lie within the loops predicted
by the BD simulations.

Remarkably,  HI appears to have almost no influence on
stress-configuration hysteresis. Although the maximum values of
the stress and $\Delta n^\ast$ are larger when free-draining
models are used, the simulations' and approximations' predictions
with and without HI show significant overlap. Similar behaviour is
also observed on a plot of $\NIps$-versus-$\Resqs$
\citep{prabhakar:phd-05}. The reason behind this insensitivity of
stress-conformational hysteretic behaviour to HI is not
understood.

\subsubsection{\label{s:refb} Steady extensional flow}

\begin{figure}[t]
\centerline
{\resizebox{7cm}{!}{\includegraphics{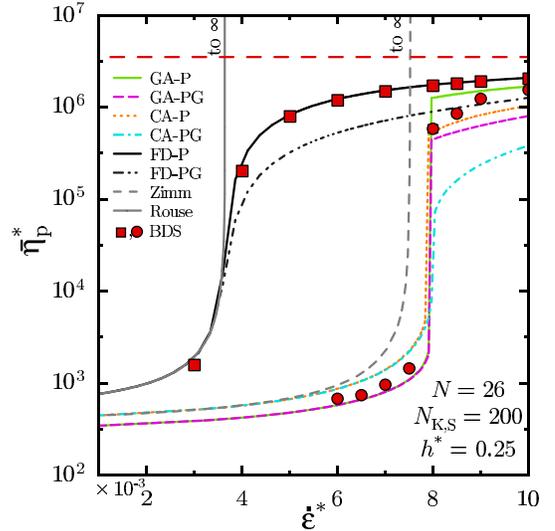}}}
\caption{\label{f:sueta} Variation of the steady-state polymer
extensional viscosity with extension-rate, for FEBS chains with HI
with and without HI. The red square and circle symbols represent
results obtained with BDS of FEBS chains with and without HI
respectively. The horizontal dashed red line is the dimensionless
extensional viscosity of fully-stretched free-draining chains
calculated using Eq.~\ref{e:fdfseta}, and marks the upper limit of
the extensional viscosity for FEBS chains with parameter values
shown.}
\end{figure}

Figure~\ref{f:sueta} compares the predictions of the steady-state
polymer extensional viscosity by different approximations with the
simulations' results obtained for FEBS chains with and without HI.
The BDS results were obtained using step changes in the extension
rates and integrating until a steady value is obtained for
$\etaps$ at each extension rate. The most important characteristic
of the curves for the steady-state extensional viscosity is the
occurrence of a sudden rapid increase in its value within a narrow
range of strain rates around a critical elongation rate $\edotsc$.
As is well known, the value of $\edotsc$ can be calculated
analytically in the Rouse model to be $\edotsc = 0.5/
\tau_1^{\ast\,\textsc{r}} = \sin^2 (\pi/2N)$ \citep{bird:dpl2} at
which the steady-state $\eetaps$ predicted by the Rouse model
diverges.  A similar divergence is predicted with the Zimm model
at $\edotsc = 0.5/ \tau_1^{\ast\,\textsc{z}}$. Here,
$\tau_1^{\ast\,\textsc{r}}$ and $\tau_1^{\ast\,\textsc{z}}$ are
the largest time constants is the respective spectra of relaxation
times in the Rouse and Zimm models. For FEBS chains with or
without HI, $\eetaps$ remains bounded by the limit for
fully-stretched free-draining chains completely aligned in the
direction of extension \citep{hassager:jcp-74},
 \begin{gather}
\eetaps = \NKS \NS (\NS +1) (\NS+2)\,. \label{e:fdfseta}
\end{gather}

As observed earlier in shear flows, the steady-state
$\eetaps$-versus-$\edots$ curves predicted by models with FEBS
chains are indistinguishable from those obtained with Rouse chains
until FE becomes important. Predictions with the FD-P
approximation in the super-critical regime ($\edots > \edotsc$)
are in remarkable agreement with the BDS results, whereas the
FD-PG approximation significantly underpredicts the exact results.
The accuracy of the FD-P approximation in its prediction of the
steady-state extensional viscosity is well known
\citep{brule:jnnfm-93}. This behaviour however is in contrast with
that observed in steady shear flows (Fig.~\ref{f:fdsspsi1}), where
the FD-PG approximation's prediction are closer to the
simulations' results even at high shear rates.

As mentioned earlier, Lielens \emph{et al.\
}\citep{lielens:jnnfm-98} and Wiest\citep{wiest:jor-89} have
argued that the accuracy of the FENE-P approximation at high
extension rates stems from the fact that the exact distribution of
spring lengths is very narrow and close to being a
$\delta$-function centered at some value $\overline{Q}$ lesser
than $Q_0$. Under such conditions, the FENE nonlinearity $\xi$ is
well approximated by the FENE-P function $\overline{\xi}$. When
$\overline{Q}$ is not very close to $Q_0$, it is clear that the
additional term in the FENE-PG function $\bm{L}$ in
Eq.~\eqref{e:Ldef} leads to inaccuracies.

We recall that higher order terms in the series expansion in
Eq.~\eqref{e:feapx} have been dropped in obtaining $\bm{L}$ in
Eq.~\eqref{e:Ldef}. Since every term [Eq.~\eqref{e:feserterm}] in
the series in Eq.~\eqref{e:feapx} is positive, it is evident that
\emph{after} applying the Peterlin approximation to every term in
the series, the full summation (if convergent) will lead to an
even greater overestimation of the influence of spring-force
fluctuations when the true spring length distribution is narrow.
Since the series expansion in Eq.~\eqref{e:feapx} itself has been
derived under the assumption of a Gaussian distribution, it
appears that the extension of the Gaussian approximation to the
FENE springs may not be appropriate in situations where the spring
length distribution is narrow, but $\overline{Q}$ is not close to
$Q_0$. The better agreement of predictions with FENE-PG springs
with BDS data observed earlier in shear flows not only indicates
that spring length fluctuations are larger in shear flows, but
also shows that the Gaussian approximation captures the influence
of these fluctuations more accurately. As the strain-rate
increases and $\overline{Q}$ approaches $Q_0$, it was shown in
section~\ref{s:fenepg} that the additional contribution in the
FENE-PG spring force becomes negligible in comparison with
$\overline{\xi}$, and the differences between the predictions of
the two approximations become smaller.

For chains with HI, the superiority of the Gaussian approximation
is clearly evident in Fig.~\ref{f:sueta}, in the sub-critical
regime. At strain-rates greater than $\edotsc$, it is the GA-P
model that performs best (also see Fig.~\ref{f:suetamult} below).
This is somewhat surprising, since one might expect from the
better performance of the FD-P model for free-draining chains
observed earlier, that fluctuations are small when chains are
stretched out in the super-critical regime. Therefore, if the
influence of both spring-force and HI fluctuations are
unimportant, the predictions of the CA-P model should be close to
the BDS data. Instead, however, the GA-P model does much better
than the CA-P model in Fig.~\ref{f:sueta}. One plausible
explanation for this observation could be that although
fluctuations in spring lengths may be small and the FENE-P
approximation for the spring force is accurate, the distribution
of overall chain configurations may not be as sharply peaked at
strain rates above $\edotsc$. The resultant fluctuations in HI may
therefore may still play an important role. The GA-PG model
underpredicts the BDS results for the extensional viscosity when
$\edots > \edotsc$, and this is seen to be the result of the
inaccuracy of the FENE-PG spring force at high extensions.

\begin{figure}[!t]
\centerline {\resizebox{7cm}{!}{\includegraphics{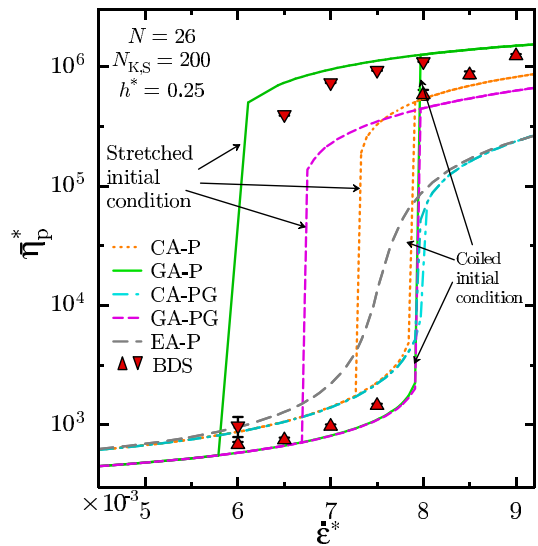}}}
\caption{\label{f:suetamult} Prediction of coil-stretch hysteresis
in the steady-state extensional viscosity by closure
approximations for FEBS chains with HI. The upright and inverted
triangle symbols represent BDS data obtained by successively
stepping up, and stepping down, the extension rate, respectively.}
\end{figure}

It is observed that the approximations for FEBS chains with HI can
predict multiple steady-state values for the extensional viscosity
for a range of values of the extension rate $\edots$
[Fig.~\ref{f:suetamult}]. The variation of the steady-state
extensional viscosity with extension-rate predicted by the
approximations in Fig.~\ref{f:suetamult} was obtained as follows.
Starting with the equilibrium second moments ($\bsigs_{ij, \,
\textrm{eq}} = \delta_{ij} \ut$) as the initial guess, the set of
coupled, nonlinear equations for the steady-state second moments
were first iteratively solved at a small value ($\ll \edotsc$) of
the extension rate. After calculating the extensional viscosity
with the converged solution for the equilibrium second moments,
this solution is used as the guess for obtaining the solution
numerically at a slightly larger value of $\edots$. This process
is continued until $\edots$ is well beyond the value of $\edotsc$
for the coil-to-stretch transition. The $\eetaps$-versus-$\edots$
curve obtained thus is indicated as having been obtained with a
``coiled initial condition" in Fig.~\ref{f:suetamult}. The process
is then reversed by successively lowering the extension-rate. In
Fig.~\ref{f:suetamult}, it is seen that with the notable exception
of the EA-P model, the predicted $\eetaps$-versus-$\edots$ curve
follows a different route as $\edots$ is decreased for all the
other approximations for chains with HI. Since this steady-state
solution branch is obtained starting from a highly ``stretched
initial condition", it is indicated as such in
Fig.~\ref{f:suetamult}. A similar procedure described earlier by
Schroeder \emph{et al.\ }\citep{schroeder:macro-04} was used to
obtain the BDS data.

The existence of such ``coil-stretch hysteresis" was first
proposed in a landmark study by De Gennes \citep{degennes:jcp-74},
who used a variable friction coefficient in a simple dumbbell
model to model the effect of the deformation of the polymer coil
on its hydrodynamics. A further closure approximation equivalent
to the consistent-averaging treatment was used to obtain a
solvable dumbbell model that predicted coil-stretch hysteresis in
a uniaxial extensional flow. To the best of our knowledge,
Fig.~\ref{f:suetamult} is the first demonstration that closure
approximations for FEBS chain models with HI can predict
coil-stretch hysteresis.

It is interesting to note that the size of the window of strain
rates where multiple steady-states are obtained, depends strongly
on the approximation used. The window size appears to depend on
the difference between the extensional viscosities in the coiled
state (the lower branch of the hysteresis curve) and the stretched
state (the upper branch of the hysteresis curve). Fluctuations in
HI lead to a widening of the window because these fluctuations
decrease $\eetaps$ in the coiled state, but increase $\eetaps$
when the chains are highly stretched. As before, the BDS data are
quite close to the predictions of the GA-P model. Significantly,
multiple steady-states are never obtained with the EA-P model, and
in the free-draining models. As pointed out by
\citeauthor{degennes:jcp-74}, coil-stretch hysteresis occurs
principally because of the modification of the mobility matrix due
to the deformation of the polymer coil. With equilibrium averaged
HI (and in all free-draining models), the mobility matrix is
deformation-independent, and hence no coil-stretch hysteresis is
observed.

As mentioned above, De Gennes first obtained such multiple
steady-states using a mean-field closure approximation in a simple
dumbbell model with a variable configuration-dependent drag
coefficient. Later, Fan \emph{et al.\ }\citep{fan:jnnfm-85}
obtained the solution to the Fokker-Planck equation for the
dumbbell model numerically without using any closure
approximations, and found that the resulting steady-state
expectations are single-valued functions of the extension rate.
They argued that De Gennes' prediction of a coil-stretch
hysteresis must therefore be an artifice  of the mean-field
approximations used in his model. More recently, Schroeder
\emph{et al.\ }\citep{schroeder:science-03} showed that the exact
steady-state configurational distribution in the variable drag
dumbbell model is bi-modal, and that a large (effective) free
energy barrier between the coiled and stretched states leads to
ergodicity-breaking. In contrast, however, the use of closure
approximations leads to two distinct uni-modal Gaussian solutions
to the modified Fokker-Planck equations.

The BDS results obtained by Schroeder \emph{et al.\
}\citep{schroeder:science-03}, and those in Fig.~\ref{f:suetamult}
suggest that the exact distribution of the steady-state
extensional stress in FEBS chains with HI is also bi-modal. The
linearity of the Fokker-Planck equation [Eq.~\eqref{e:FPEconn}] in
the configurational probability distribution function further
guarantees that its exact solution is unique. The introduction of
any mean-field approximation however leads to a modified
Fokker-Planck equation that is nonlinear in the probability
distribution function, since the coefficients in the modified
Fokker-Planck equation are functionals of the distribution
function. While the solution of this modified equation may be a
uni-modal Gaussian distribution, the nonlinearity in the
distribution function means that the solution may no longer be
unique. Thus, the exchange of the nonlinearities due to FENE and
HI for a nonlinearity in the distribution function destroys the
multi-modal nature of the original solution, but allows the
modified Fokker-Planck equation to have multiple solutions.

In spite of the fundamentally different probability distributions,
the results presented in Fig.~\ref{f:suetamult} above indicate
that the multiple steady-states  obtained with the approximations
can closely follow the long-lived kinetically trapped states in
the original model. Therefore, closure approximations for FEBS
chains with HI may prove to be useful in a detailed exploration of
hysteretic phenomena  caused by ergodicity-breaking in dilute
polymer solutions.

We next briefly discuss the accuracy of the diagonalization
approximations in shear and extensional flows.

\subsection{\label{s:tfn} Two-fold Normal Approximation}

\begin{figure}[t]
\centerline {\resizebox{7cm}{!}{\includegraphics{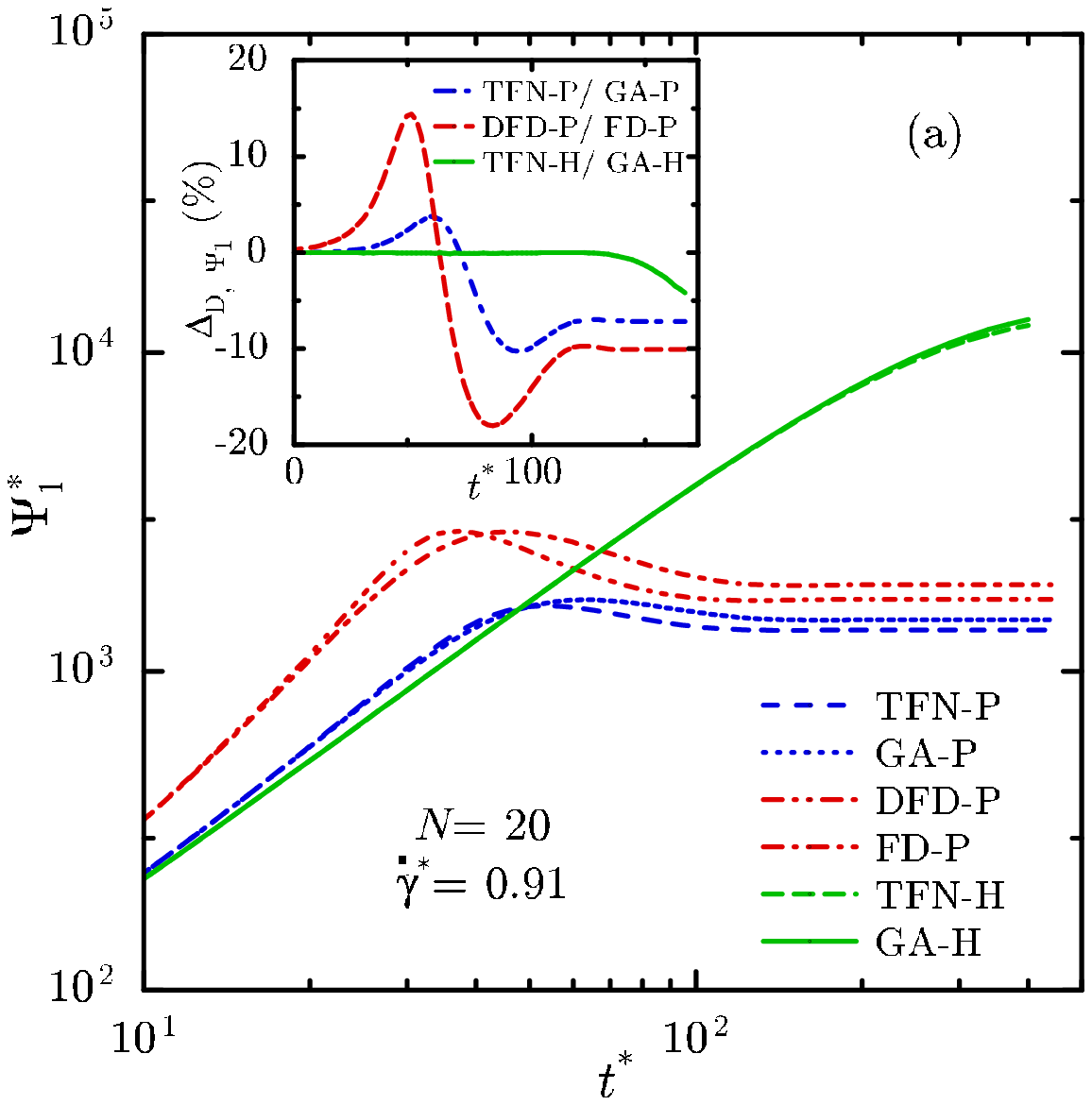}}}
\centerline {\resizebox{7cm}{!}{\includegraphics{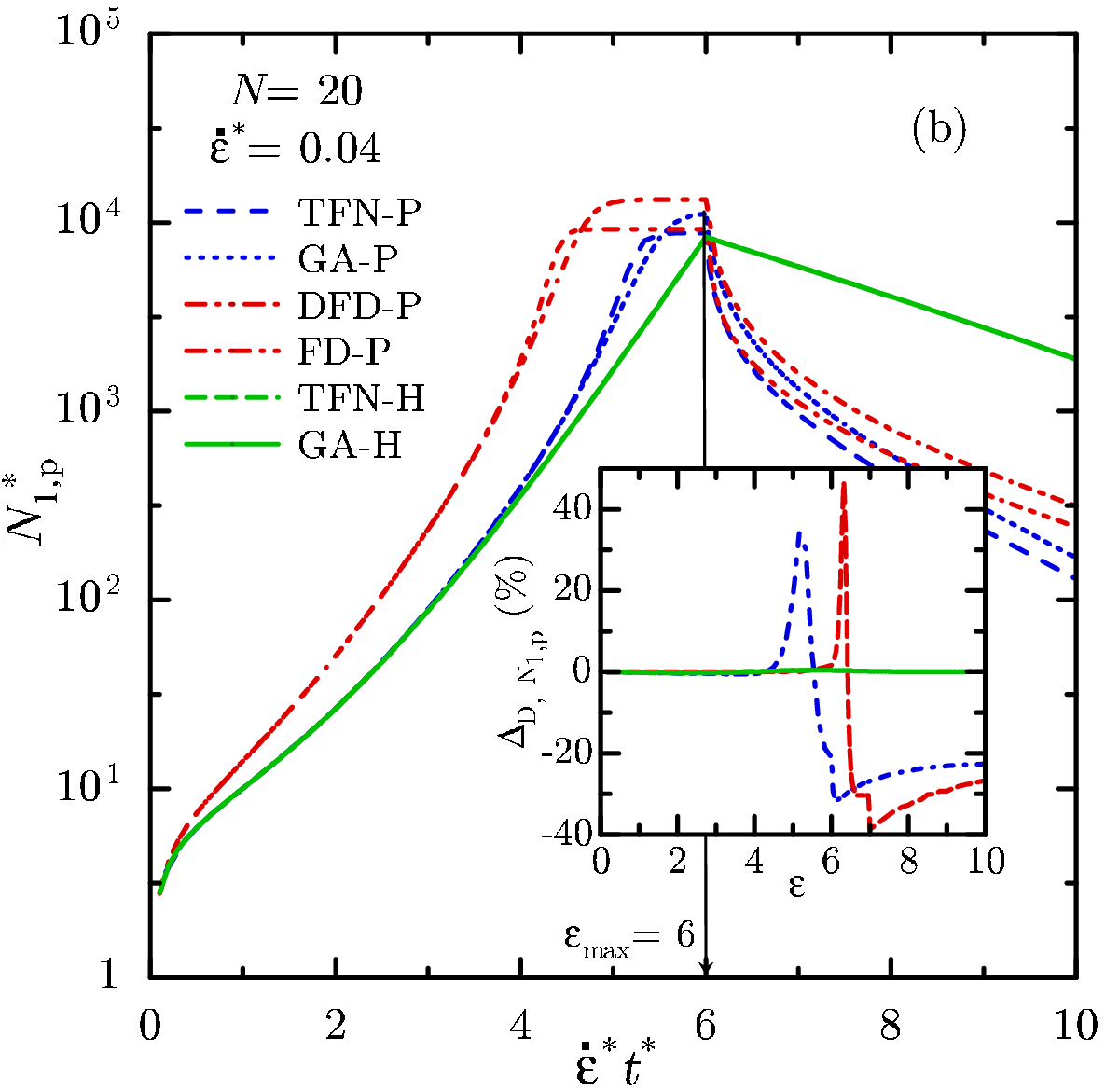}}}
\caption{\label{f:uspsi1diag} Predictions of the diagonalization
approximations for (a) the growth of the first normal-stress
difference coefficient after imposition of a steady shear rate,
and (b) the growth and relaxation of the polymer contribution to
the dimensionless first normal-stress difference, during start-up
and following cessation, of steady extensional flow. In (a), $\NKS
= 18.3$ for models with FENE-P springs, and $\hs = 0.25$ for
models with HI. In (b), $\NKS = 50$ for models with FENE-P
springs, and $\hs = 0.25$ for models with HI. The insets plot the
relative deviations of the predictions of the diagonalized and
un-diagonalized approximations, calculated using
Eq.~\eqref{e:diagdev}.}
\end{figure}

Figure~\ref{f:uspsi1diag} compares the predictions at high shear
and extension rates  obtained with the diagonalized DGA-H, DFD-P
and TFN-P approximations and their respective un-diagonalized
versions, namely, the GA-H, FD-P and GA-P approximations. The
effect of making the diagonalization assumption is examined here
by calculating the relative deviation between the predictions of a
property $\pi$ obtained with an approximation ``A" and its
diagonalized version ``DA":
\begin{gather}
\Delta_{\textsc{d},\,\pi} \equiv \frac{\pi_\textsc{da} -
\pi_\textsc{a}}{\pi_\textsc{a}} \,. \label{e:diagdev}
\end{gather}
Figures~\ref{f:uspsi1diag}~(a) and~(b) demonstrate that the
relative deviation between the predictions of GA-P model and its
diagonalized version in shear flow, the TFN-P model, can be quite
significant. In contrast, the diagonalization assumption is much
more accurate for approximations for Rouse chains with HI (the
GA-H and TFN-H models), for which the largest relative deviation
does not exceed 10\% in Fig.~\ref{f:uspsi1diag}~(a) and (b). On
the other hand, the overall time variation of the deviation
between the TFN-P and GA-P models is qualitatively similar to that
observed between the diagonalized FD-P (DFD-P) model and the FD-P
model, but is smaller in magnitude. The magnitude of the
deviations depend on the property considered. In shear flows, the
deviations in the predictions for $\etaps$ are smaller. In both
shear and extensional flows, the deviations are larger in the
predictions of $\Resqs$ than in the stresses.

The results above indicate that for Rouse chains, or for FEBS
chains at low strains and/ or strain-rate, when FE is unimportant,
the inclusion of HI diminishes the influence of the off-diagonal
components of the normal-mode covariances on the overall behaviour
predicted by the models, and the diagonalization approximations
are quite accurate. The diagonalization assumption begins to break
down when FE begins to exert its influence.

\begin{table}[!t]
\caption{\label{t:cput} Sample comparison of CPU-time requirements
for BD simulations, un-diagonalized and diagonalized
approximations. The value reported for BD simulations is for
$10^4$ trajectories. For $N = 160$, the reported value has been
calculated from the CPU-time required for 200 trajectories. The
CPU-times reported for the CA-P and GA-P approximations for $N =
160$ have been projected from the CPU-time--versus--$N$ scaling
observed for $20 \leq N \leq 100$. All calculations were performed
on 2.66 GHz, Pentium 4 (Dell Precision 250) processors.}
\addtocounter{table}{-1}
\begin{longtable}[c]{p{0.7in} p{0.7in} p{0.7in} p{0.7in} p{0.7in}}
\hline \hline
$N$ & \multicolumn{4}{c}{CPU-time in hrs.} \\
 & BDS  & CA-P & GA-P & TFN-P \\
 \hline
 60 & 220 & 10 & 51 & 0.9 \\
 160  & 10012  & 2073   & 3131  & 32 \\
 \hline
 \hline
\end{longtable}
\end{table}

In spite of the deterioration in the quality of the
diagonalization approximations when FE is important, the massive
reduction in CPU-time usage (Table~\ref{t:cput}) achieved with
these approximations make them an ideal tool for rapid preliminary
exploration of vast regions of parameter space. In fact,
Eq.~\eqref{e:tfnpqq} coupled with the equation for the polymer
stress in Eq.~\eqref{e:tauGA}, constitute a new constitutive model
for the polymer stress in dilute polymer solutions. This
constitutive model reproduces all the key features characteristic
of these solutions with reasonable accuracy, including
coil-stretch hysteresis (not shown here), and thus represents a
significant improvement over constitutive equations such as the
Oldroyd-B or FENE-P dumbbell equations typically used in the
simulation of complex flows of dilute polymer solutions.

\section{\label{s:concl} Summary and Conclusions}

The Gaussian Approximation was originally suggested for Hookean
dumbbells and Rouse chains with HI by
\"{O}ttinger\citep{ottinger:jcp-89} and
Wedgewood\citep{wedgewood:jnnfm-89}, and has been shown to improve
predictions by approximately accounting for the influence of
fluctuations in HI. In this work, we use the same idea in the
context of a free-draining bead-FENE spring model to first obtain
a new approximation for tackling the closure problem posed by the
FENE nonlinearity. We show that the resulting equations in the
approximate model can be interpreted in terms of a mean-field
spring potential, and that this ``FENE-PG" spring potential
improves upon the well known FENE-P approximation in situations
where spring force fluctuations are important. Furthermore, the
use of additional diagonalization-of-normal-modes assumptions for
simplifying the model equations lead to considerable savings in
computation time, without significant loss in accuracy.

The direct application of the Gaussian Approximation for
bead-spring models simultaneously incorporating the effects of
both FE and HI leads to an undesirable dependence of the
predictions of equilibrium static properties on the hydrodynamic
interaction parameter, $\hs$. Instead, we use a two-step procedure
to marry the FENE-PG mean-field spring potential with the Gaussian
Approximation for HI to arrive at a closed set of equations for
the second-moments, which is consistent with the equilibrium
Boltzmann probability distribution. This approximate model
accounts for the influence of fluctuations in both HI and the
spring forces.

The method of combining the Gaussian approximation for HI with the
FENE-PG approximation for HI suggested in this study is found to
lead to predictions that compare well with the results of exact BD
simulations for properties in steady and unsteady shear flows. The
better agreement of the predictions of this approximate
model---dubbed here as the ``GA-PG" model---than those obtained
with the FENE-P and/or Consistent-Averaging approximations
emphasizes the importance of fluctuations in HI, and also
highlights the role spring force fluctuations play at moderate and
high shear rates. In extensional flows, the GA-P model combining
the Gaussian approximation for HI with FENE-P springs is found to
perform better when chains are highly stretched, perhaps
indicating that under such conditions, fluctuations in HI may
still be important, even though local fluctuations in spring
lengths are small.

The predictions obtained with the Gaussian approximation for HI in
steady and unsteady shear and extensional flows confirms previous
observations on the role of fluctuations in HI on macroscopic
behaviour.  When chains are close to their equilibrium isotropic
states, fluctuations in HI enhance the screening of the solvent
velocity gradient caused by HI. The role of fluctuations in HI
reverses in situations where polymer chains experience significant
stretching, and aid the penetration of the velocity field into the
stretched polymer coil. Consequently, the solvent is able to
engage a firmer ``grip" on the chains and this tends to deform an
already anisotropic chain further. Fluctuations in the FENE spring
force are found to increase the chains' resistance to stretching.

When the stress predicted in BD simulations of FEBS chains during
the start-up and following the cessation of a strong extensional
flow is plotted against predictions of the mean-squared end-to-end
distance of the chains, or against the intrinsic birefringence,
large hysteresis loops were observed. The hysteresis loops
predicted with closure approximations were much smaller than those
predicted with the simulations. Surprisingly, HI was found to have
a negligible influence on the phenomenon of stress-conformational
hysteresis observed in extensional flows.

Closure approximations for FEBS chains with HI predict
coil-stretch hysteresis. In particular, predictions of hysteresis
in the extensional viscosity by the GA-P model are in excellent
agreement with the exact results obtained with BD simulations. It
is observed that fluctuations in HI enhance the hysteretic effect,
increasing the width of the coil-stretch hysteresis window on a
plot of the steady-state extensional viscosity versus the
strain-rate.

\appendix

\section{\label{s:appA} Tensors associated with hydrodynamic interactions }

The tensor $\bAb_{ij}$ in Eq.~\eqref{e:gapgqq}, which describes
contributions arising from averaging the Oseen-Burgers tensor with
the non-equilibrium distribution function, is given by the
expression,
\begin{align}
\begin{split}
\bAb_{ij} \equiv \langle \bA_{ij} \rangle =  A_{ij} \ut &+
\sqrt{2} \hs \left[ \bm{H}_{ij} + \bm{H}_{i+1,\,j+1} \right. \\&-
\left. \bm{H}_{i+1,\,j} - \bm{H}_{i,\,j+1} \right]\,, \end{split}
\label{e:babij}
\end{align}
while the tensors $\bm{\Delta}_{ij}$ and
$\bm{\Delta}_{ij}^\textsc{t}$, which represent contributions due
to fluctuations in HI, are given by,
\begin{gather}
\bm{\Delta}_{ij} \equiv \sum_{r,s = 1}^{\NS}  \bGam^{is}_{rj}
\colon
(\bsigs_{sr} \cdot \bm{L}_r), \notag \\
\intertext{and,} \bm{\Delta}_{ij}^\textsc{t} = \sum_{r,s =
1}^{\NS} (\bm{L}_r \cdot \bsigs_{rs}) \colon  \bGam^{si}_{jr} \,,
\label{e:deltaij}
\end{gather}
with,
\begin{align}
\begin{split}
\bGam^{ps}_{rq} \equiv \frac{3 \sqrt{2} \hs}{4} [ &
\DT{r}{q}{p}{s} \bm{K}_{rq} + \DT{r+1,}{\,q+1}{p}{s}
\bm{K}_{r+1,\,q+1} \\& - \DT{r+1,}{\,q}{p}{s}\bm{K}_{r+1,\,q} -
\DT{r,}{\,q+1}{p}{s}\bm{K}_{r,\,q+1}]\,.
\end{split}\label{e:bgampsrq}
\end{align}
The second-rank tensorial function $\bm{H}_{\nu \mu} = \bm{H}
(\bm{S}_{\mu \nu}^\ast )$ is related to the average of the HI
tensor through,
\begin{align}
\langle \zeta \bO_{\nu \mu}  \rangle &= \sqrt{2} \hs \bm{H}
(\bm{S}_{\mu \nu}^\ast )\,,
\end{align}
whereas the fourth-rank tensorial function  $\bm{K}_{\mu \nu} =
\bm{K}(\bm{S}^\ast_{\mu \nu} )$ is related to the HI tensors
through,
\begin{gather}
\langle\frac{ \partial ( \zeta \bO_{\nu \mu})  }{\partial \br_{\mu
\nu}} \br_{\mu \nu} \rangle = \frac{3 \sqrt{2} \hs }{4} \bm{K}
(\bm{S}^\ast_{\mu \nu} ) \cdot \bm{S}^\ast_{\mu \nu} \, .
\label{e:Kdef2}
\end{gather}
In the equations above,
\begin{gather}
\bm{S}^\ast_{\mu \nu} = \langle \br^\ast_{\mu \nu} \br^\ast_{\mu
\nu} \rangle = \sum_{i,j = 1}^{\NS}\DT{\mu}{\nu}{i}{j}
\bsigs_{ij}\,. \label{e:Sneq}
\end{gather}
Here,
\begin{gather}
\DT{\mu}{\nu}{i}{} =
\begin{cases}
1, & \text{ if } \mu \leq i < \nu \, , \\
-1, & \text{ if } \nu \leq i < \mu \, , \\
0 \, , &\text{ otherwise,}
\end{cases}
\end{gather}
is the one-dimensional ``box" function. This function can be
succinctly expressed in terms of the Heaviside step function,
\begin{gather}
\Theta (m, n) =
\begin{cases}
1, & \text{ if } n \geq m \\
0 & \text{ otherwise },
\end{cases}
\end{gather}
as $ \DT{\mu}{\nu}{i}{} = \Theta (\mu, i) - \Theta (\nu, i) = -
\DT{\nu}{\mu}{i}{}$. Further, the dyad
\begin{gather}
\br_{\mu \nu} \br_{\mu \nu} = \sum_{i, j \, = \min(\nu,
\mu)}^{\max (\nu, \mu) - 1} \bQ_i \bQ_i = \sum_{i, j \, = 1}^{\NS}
\DT{\mu}{\nu}{i}{j} \bQ_i \bQ_j \,,
\end{gather}
where the two-dimensional box function,
\begin{gather}
\DT{\mu}{\nu}{i}{j} = \DT{\mu}{\nu}{i}{} \DT{\mu}{\nu}{j}{} \,.
\label{e:2dbox}
\end{gather}
This function is symmetric with respect to an exchange of the
indices $\mu$ and $\nu$, and/or $i$ and $j$.

When the Oseen-Burgers form of the HI tensor is used, it can be
shown that \citep{ottinger:jcp-87c, zylka:macromol-91}
\begin{gather}
\bm{H}_{\nu \mu} = \frac{3}{2 (2 \pi)^{3/2}} \int  \,
\frac{1}{k^2} \left( \ut - \frac{\bm{k} \bm{k}}{k^2} \right) \exp
( -\half\, \bm{k} \bm{k} \colon \bm{S}_{\mu \nu}^\ast) \, d
\bm{k}\,, \label{e:Hfndef}
\end{gather}
and
\begin{align}
\bm{K}_{\nu \mu} = \frac{-2}{(2 \pi)^{3/2}} \int \, \frac{1}{k^2}
\bm{k} \left( \ut - \frac{\bm{k} \bm{k}}{k^2} \right) \bm{k}
\,\exp ( -\half\, \bm{k} \bm{k} \colon \bm{S}_{\mu \nu}^\ast ) \,d
\bm{k} \,, \label{e:Kdef}
\end{align}
Thus, the functions $\bm{H}_{\nu \mu}$ and $\bm{K}_{\nu \mu}$, and
therefore $\bAb_{ij}$ and $\bm{\Delta}_{ij}$ are completely
expressible in terms of the second moments of the Gaussian
distribution.

The inner product of $\bm{K}$ with a second-rank tensor
$\mathbf{a}$ is defined in component form as
\begin{gather}
(\bm{K}\colon \mathbf{a} )_{\alpha \beta} = \sum_{ \gamma
,\delta\, = 1}^{3} K_{\alpha \beta \gamma
\delta} \, a_{ \delta \gamma},\\
\intertext{and,} (\mathbf{a} \colon \bm{K} )_{\alpha \beta} =
\sum_{ \gamma ,\delta \, = 1}^{3} a_{ \delta \gamma}\,K_{\gamma
\delta \alpha \beta }\,\text{ for $\alpha$, $\beta$ = 1,2,3 .}
\end{gather}
Further, since the inner product
\begin{gather*}
\frac{1}{k^2} \left( \ut - \frac{\bm{k} \bm{k}}{k^2} \right)
\bm{k} \colon \ut = \bm{0} \,,
\end{gather*}
for all $\bm{k}$, we have
\begin{gather}
\bm{K} \colon \ut = \bm{0} \,.
\end{gather}
A direct consequence of this result is that the $\bm{\Delta}$
tensors defined through Eqs.~\eqref{e:deltaij} and
\eqref{e:bgampsrq}, vanish under isotropic equilibrium conditions.


\bibliographystyle{unsrtnat}

\end{document}